\documentclass[prx,twocolumn,showpacs,amsfonts,amsmath,floatfix]{revtex4-2}

\usepackage{amsmath, amsfonts, amssymb, amsthm, bbm}
\usepackage{dsfont}
\usepackage[T1]{fontenc}
\usepackage{enumitem}
\usepackage{bm}
\usepackage{mathtools}
\usepackage{graphicx}
\usepackage{epstopdf} 
\usepackage{epsfig}
\usepackage{dcolumn}
\usepackage{bm}
\usepackage{hyperref}
\usepackage{times}
\usepackage{physics}
\usepackage{graphicx}
\usepackage{booktabs}
\usepackage{epsfig}
\usepackage{listings}
\usepackage[table,xcdraw]{xcolor}
\usepackage{multirow}
\usepackage{colortbl}
\input{Qcircuit}
\makeatletter

    \def\CT@@do@color{%
      \global\let\CT@do@color\relax
            \@tempdima\wd\z@
            \advance\@tempdima\@tempdimb
            \advance\@tempdima\@tempdimc
    \advance\@tempdimb\tabcolsep
    \advance\@tempdimc\tabcolsep
    \advance\@tempdima2\tabcolsep
            \kern-\@tempdimb
            \leaders\vrule
                    \hskip\@tempdima\@plus  1fill
            \kern-\@tempdimc
            \hskip-\wd\z@ \@plus -1fill }
    \makeatother

\definecolor{orange}{HTML}{FF8900}
\definecolor{ceruleanblue}{rgb}{0.16, 0.32, 0.75}

\newcommand{\logic}[1]{{\bm{#1}}} 
\newcommand{\logict}[1]{{\textbf{#1}}} 

\begin{document}

\title{Efficient simulatability of continuous-variable circuits with large Wigner negativity}

\author{Laura Garc\'{i}a-\'{A}lvarez$^{1}$, Cameron Calcluth$^1$, Alessandro Ferraro$^{2}$, Giulia Ferrini$^{1}$}
\affiliation{$^1$ Department of Microtechnology and Nanoscience (MC2), Chalmers University of Technology, SE-412 96 G\"{o}teborg, Sweden \\
$^2$ Centre for Theoretical Atomic, Molecular and Optical Physics, Queen's University Belfast, Belfast BT7 1NN, United Kingdom
}


\begin{abstract}
 Discriminating between quantum computing architectures that can provide quantum advantage from those that cannot is of crucial importance.
 From the fundamental point of view, establishing such a boundary is akin to pinpointing the resources for quantum advantage; from the technological point of view, it is essential for the design of non-trivial quantum computing architectures.
 Wigner negativity is known to be a necessary resource for computational advantage in several quantum-computing architectures, including those based on continuous variables (CVs).
 However, it is not a sufficient resource, and it is an open question under which conditions CV circuits displaying Wigner negativity offer the potential for quantum advantage. In this work we identify vast families of circuits that display large, possibly unbounded, Wigner negativity, and yet are classically efficiently simulatable, although they are not recognized as such by previously available theorems.
 These families of circuits employ bosonic codes based on either translational or rotational symmetries (\textit{e.g.}, Gottesman-Kitaev-Preskill or cat codes), and can include both Gaussian and non-Gaussian gates and measurements. Crucially, within these encodings, the computational basis states are described by intrinsically negative Wigner functions, even though they are stabilizer states if considered as codewords belonging to a finite-dimensional Hilbert space. We derive our results by establishing a link between the simulatability of high-dimensional discrete-variable quantum circuits and bosonic codes.
\end{abstract}


\maketitle


\section{Introduction}
\label{sec:intro}
With the advent of Noisy Intermediate-Scale Quantum Computing (NISQ) devices~\cite{preskill2018} and quantum computational advantage~\cite{arute2019,wang2019}, it becomes of paramount importance to identify resourceful architectures --- \textit{i.e.}, those that are capable of yielding quantum speed-up for computation \cite{nielsen2000}--- and to distinguish them from those that cannot.
A paradigmatic example of a criterion that serves this purpose for discrete-variable (DV) architectures is provided by the Gottesman-Knill theorem~\cite{gottesman1999}. The latter states that all architectures composed of stabilizer input states (\textit{i.e.}, eigenstates of the Pauli operators), Clifford operations (\textit{i.e.}, unitary operations that map the Pauli group to the Pauli group via conjugation), and Pauli measurements, can be efficiently simulated with classical computers and therefore they are computationally resourceless.

An alternative to DV approaches is constituted by CV architectures~ \cite{braunstein:05, weedbrook2012, serafini2017}, which are especially promising in terms of scalability and
fault tolerance~\cite{takeda2019, pfister2019continuous}.
In particular, in the optical regime, scalable architectures have been demonstrated, where more than one million optical modes have been entangled~\cite{yoshikawa2016invited} and universal two-dimensional cluster states have been generated~\cite{asavanant2019, asavanant2019a, larsen2019deterministic}.
Furthermore, within the microwave regime and with the crucial use of \textit{bosonic codes} \cite{chuang1997}, progresses towards error correction and fault tolerance have also been recently made. Remarkably, logical qubits encoded in microwave field resonators have been shown to exceed the lifetime of their physical constituents~\cite{ofek2016} --- with further improvements within reach~\cite{romanenko2020}.
The landmark achievement in Ref.~\cite{ofek2016} made use of the so-called cat encoding \cite{mirrahimi2014}, a specific type of rotation-symmetric bosonic (RSB) code~\cite{grimsmo2020}, where logical qubit states are encoded in superpositions of coherent states with opposite phase.
Recently, another kind of bosonic code capable of promoting CV architectures to fault tolerance --- the Gottesman-Kitaev-Preskill (GKP) code~\cite{gottesman2001}, based on translational invariant states --- has also been demonstrated experimentally with both microwave technology~\cite{campagne-ibarcq2019} and trapped ions~\cite{fluhmann2018,fluhmann2019}.

So far, the most used criteria for setting a boundary to the power of CV quantum computing architectures, in analogy to the Gottesman-Knill theorem, are based on the positivity of quasi-probability distributions~\cite{rahimi-keshari2016}, and in particular of the Wigner function: It always exists a classical algorithm that can simulate efficiently the output of a quantum circuit with input states, unitary operations, and measurements described by non-negative Wigner functions~\cite{mari2012,veitch2013}.
Unfortunately, in the framework of bosonic codes, this criterion is intrinsically of little use. This is the case since, in general, basis states for codes --- such as the GKP or RSB codes --- need to be orthogonal, which in turn implies that they must display negativities in their Wigner functions~\footnote{Pure states with non-negative Wigner function can only have Gaussian Wigner functions~\cite{hudson1974}. Therefore, in general, no two states of this type can be orthogonal.}.
Therefore, no CV circuit that uses bosonic codes to process information can be assessed using the aforementioned criteria, regardless of whether it can provide quantum computational advantage or not.

To exemplify this impasse, consider a CV architecture composed of initial stabilizer GKP-encoded states, over which encoded Clifford operations and computational basis measurements act. These circuits are clearly classically efficiently simulatable, as recognized by the standard Gottesman-Knill theorem for qubits, since we can give an effective discrete-variable interpretation of the architecture. However, the known CV criteria fail to recognize this, due to the Wigner negativity of the input states.
The crucial question that we ask in this paper is the following: what happens for more general circuits, such as those with a unitary evolution that does not correspond to encoded Clifford gates? Are these classically efficiently simulatable, or hard to simulate for a classical computer? For example, consider a CV circuit composed of encoded input states (therefore displaying Wigner negativities) and arbitrary Wigner-positive operations and measurements. Can we simulate it efficiently on a classical device?  To answer these questions, we cannot assess the efficient simulatability directly using algorithms for DV architectures. On the one hand, the Gottesman-Knill theorem cannot be applied since the initial encoded qubit states may be taken outside the computational space by certain operations that are not defined in the bosonic code subspace.
On the other hand, as the qubit states in bosonic codes are intrinsically Wigner negative, we cannot apply current algorithms for the simulation of CV systems, either. In other words, existing criteria are not suitable to assess the simulatability of these more general architectures.

Although we fail to answer this question for the most general types of circuits, in this paper we solve this issue for a large family of circuits, that go beyond the trivial encoding of qubits in CV and encompass instances of relevance for experimental platforms \cite{cai2017,bromley2020}. The  framework that we develop hence provides a boundary to the computational power of families of CV architectures that employ bosonic encodings, including both translation- and rotation-symmetric codes, and specific families of  circuits. Our method is based on the embedding of encoded qubit states into encoded qudit (namely, DV systems of dimension larger than two) states, where some of the CV operations that are not defined in the logical qubit space, now become Clifford operations in a higher-dimensional qudit space. We extend this formalism to also include, more generally, the encoding of systems of dimension $d_1$ into systems of dimension $d_2$, thereby accommodating for an even larger class of simulatable architectures.

Even if in existing theorems the Wigner negativity is always regarded as a necessary but not sufficient resource~\cite{bartlett2002,mari2012,veitch2013,rahimi-keshari2016}, recent work  shows that the simulation cost of general quantum computing architectures with a Monte-Carlo algorithm is  exponential in the Wigner negativity of a quantum circuit~\cite{pashayan2015,Note10}.
\footnotetext[10]{Notice that by Wigner negativity we refer to the concept of mana in DV~\cite{veitch2014}, or logarithmic Wigner negativity in CV~\cite{albarelli2016nonlinearity,takagi2018}. The mana is the logarithm of the negativity considered in Ref.~\cite{pashayan2015}.
Also notice that the notion of simulatability in Ref.~\cite{pashayan2015} refers to the estimation of a single measurement outcome. This notions of simulatibility is strictly contained in the ones considered in our paper.}
In contrast, we show that a different approach, based on a generalization of the Gottesman-Knill theorem for systems of dimension $d$, allows recognizing highly-negative Wigner function architectures as efficiently simulatable.

This work uses concepts and tools belonging to two areas of quantum information that developed largely independently: on the one hand, the research on simulatability of DV-based quantum computers and, on the other hand, the development of bosonic codes for CVs. Researchers familiar with one area are not necessarily familiar with the other. Therefore, we structured this paper in a self-consistent way. First, we state our main results in Sec.~\ref{sec:mainclaims}.
Then, in Sec.~\ref{sec:qubits-qudits}, we introduce the basic formalism for DV systems of different dimensions, and we review how to encode qubits into qudits, and more generally systems of dimension $d_1$ into systems of dimension $d_2$, following schemes developed in the context of quantum error correction.
In Sec.~\ref{sec:DV-theorems}, we review classical algorithms for simulating quantum computing DV architectures, in particular, the Gottesman-Knill theorem, and the analogous theorems for higher-dimensional systems.
In Sec.~\ref{sec:bosonic}, we recall the basics of encoding qubits in CV systems, with both translation-symmetric codes, as the GKP encoding, and rotation-symmetric or RSB codes. In the latter context, we consider an extention of the RSB codes to include higher-dimensional systems or qudits, and characterize the Clifford operations for systems of any dimension. Finally, in Sec.~\ref{sec:our-CV-theorem}, we develop our framework for assessing the simulatability of CV architecture based on embedding the logical quantum information of lower-dimensional systems into higher-dimensional systems and demonstrate our main results. We present our concluding remarks in Sec.~\ref{sec:outlook-conclusions}.

\section{Main results}
\label{sec:mainclaims}

In order to facilitate the reading, we have collected in this Section the main results of this work. The corresponding details and proofs will be given in the subsequent Sections.

In this paper, we provide a framework that recognizes as classically efficiently simulatable some families of CV architectures that include Wigner-negative input states as well as, in particular cases, Wigner-negative operations. As said, the classically efficient simulatability of Wigner-negative circuits is not {\it per se} surprising, nor does it contradict current theorems. The novelty of our framework lies in the fact that (i) it allows us to identify existing algorithms for the classical simulation of these circuits whose running times do not scale exponentially with the logarithmic Wigner negativity and (ii) it recognizes  as classically efficiently simulatable circuits of experimental relevance, especially in the context of sampling models (see Sec.~\ref{sec:implications}).

The definition and main properties of Wigner functions are recalled in Appendix~\ref{app:wigner}. We divide these families into GKP circuits and RSB circuits --- according to their input states, unitary intermediate operations, and final measurements.  The input states correspond to qudit computational basis states, which are  encoded in CV systems using either GKP or RSB codes.
The intermediate operations are related to the discrete symmetries associated to the chosen bosonic encoding (either translational or rotational symmetries) but, crucially, in a higher DV logical dimension. This allows us to vastly extend our results on simulatability from a handful set of qubit-encoded Clifford operations to operations outside the encoded qubit logical space.  Finally, the measurements considered are the encoded Pauli measurements, again in a higher DV dimension.

\begin{figure}[htpb]
 \centering
 $$
  \Qcircuit @C=2.5em @R=1.2em {
  \lstick{\ket{\logic{j_{d_1}}}}   & \qw & \multigate{2}{\mathcal C_{d_2}} & \qw & \qw & \measureD{\hat{q}} \\
  \lstick{\vdots\ \ \ \ \ }  & {/}\qw  & \ghost{\mathcal C_{d_2}} &  {/}\qw & \qw & \vdots\\
  \lstick{\ket{\logic{j_{d_1}}}} & \qw & \ghost{\mathcal C_{d_2}} &  \qw & \qw & \measureD{\hat{q}}
  }
 $$
 \caption{Efficiently simulatable  circuits encoded within the GKP framework. Input states are the $d_1$-dimensional qudit computational basis states encoded in GKP. These states are operated on by any Clifford operation existing within the $d_2$-dimensional Clifford group, with $d_2 = d_1 a^2$ for a natural number $a$.
  Homodyne measurements are performed in the position basis and correspond to Pauli $Z$ measurements in a $d_2$-dimensional system. The resulting circuit is efficiently simulatable even though \textit{(i)} the input states possess negative Wigner function (hence CV criteria for simulatability cannot be applied), and \textit{(ii)} Clifford group operations take the logical $d_1$-dimensional qudit states outside of their  codespace (hence theorems for the simulatability of $d_1$-dimensional systems cannot be applied either). In particular, we study as an example the case of initial qubit states, $d_1=2$, and subsequent qudit Clifford operations, $d_2=2 a^2$.}
 \label{fig:circuit}
\end{figure}
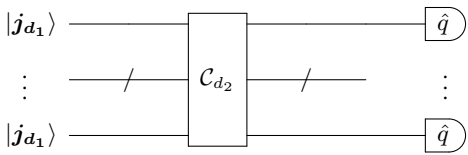

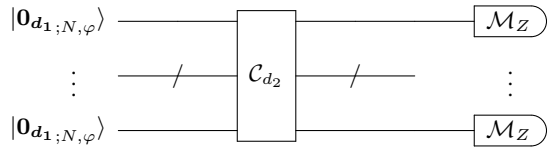
\begin{figure}[htpb]
 \centering
 $$
  \Qcircuit @C=2.5em @R=1.2em {
  \lstick{|\logic{0_{d_1}}_{;N,\varphi}\rangle}   & \qw & \multigate{2}{\mathcal C_{d_2}} & \qw & \qw & \measureD{\mathcal M_Z} \\
  \lstick{\vdots\ \ \ \ \ }  & {/}\qw  & \ghost{\mathcal C_{d_2}} &  {/}\qw & \qw & \vdots \\
  \lstick{|\logic{0_{d_1}}_{;N,\varphi}\rangle} & \qw & \ghost{\mathcal C_{d_2}} &  \qw & \qw & \measureD{\mathcal M_Z}
  }
 $$
 \caption{Efficiently simulatable stabilizer circuits encoded within the RSB framework. Input states are the $d_1$-dimensional qudit computational basis states encoded as an $N$-order rotational bosonic codeword, using the $\ket\varphi$ primitive, as defined in Eq.~(\ref{eq:codewordrbc}). These states are operated on by any Clifford operation existing within the $d_2$-dimensional Clifford group.
  The final phase measurements correspond to Pauli $Z$ measurements in the $d_2$-dimensional qudit basis~\cite{holevo2011,helstrom1969,grimsmo2020}. The resulting circuit is efficiently simulatable even though all its components (input states, gates, and measurements) display negativities in their respective Wigner functions. As in the GKP case, we study in particular the example of initial qubit states, $d_1=2$, and qudit Clifford operations, $d_2=d$. }
 \label{fig:circuitrsb}
\end{figure}

Specifically, the two families of CV architectures for GKP and RSB codes that we will show to be simulatable are sketched in Figs.~\ref{fig:circuit} and~\ref{fig:circuitrsb}, respectively. We consider an arbitrary number $n$ of CV systems, characterized by canonical position $\hat q_k$ and momentum $\hat p_k$ operators, with $[\hat q_k, \hat p_l] = i \delta_{k,l}$, and we denote with $\hat n_k$ the number operator, where $k,l=1,\dots,n$ refers to each system.
The input states are prepared in an encoded state corresponding to the logical $d_1$-dimensional computational basis --- \textit{i.e.}, the codewords $|\logic{j_{d_1}}\rangle$ and $|\logic{j_{d_1}}_{;N}\rangle$ for GKP and RSB codes, respectively, with $j=0,\dots,d_1 -1$, where the subscript refers to the dimension of the logical space, and $N$ corresponds to the order of the rotation symmetry in RSB codes (see Sec.~\ref{sec:bosonic} for their formal definitions). As said, these states have a non-positive Wigner function. Then, an arbitrary $poly(n)$ sequence of unitary operations and measurements are applied, chosen from those generated by these sets
\begin{align}
 \label{eq:GKP-operations}
  & \text{GKP} : \{e^{i\hat q_k^2/2},e^{-i\alpha\hat p_k},e^{i\alpha\hat q_k},e^{i\hat q_k\hat q_l},e^{\tfrac{i\pi}{4}(\hat p_k^2 + \hat q_k^2)},\text{HM}\},                                        \\
 \label{eq:RSB-operations-1}
  & \text{RSB} : \{e^{i\tfrac{2\pi}{ad_1 N}\hat{n}},e^{i\tfrac{\pi}{d_1}\left(\frac{\hat{n}^2}{N^{2}}-\beta\frac{\hat{n}}{a N}\right)},e^{i\tfrac{2\pi}{d_1 N^2}\hat{n}_{k}\hat{n}_{l}},\text{PM}\},
\end{align}
where $a$ is an arbitrary natural number, $\alpha = \sqrt{2\pi/d_1}/a$, and $\beta=0$ for even $d_1 a^2$ while and $\beta=1$ for $d_1 a^2$ odd. Finally, HM and PM correspond to homodyne and phase measurements~\cite{helstrom1969,holevo2011}, respectively.

For systems initialized in RSB-encoded logical states $|\logic{0_{d_1}}_{;N}\rangle$, it is possible to simulate the additional families of circuits generated by
\begin{align}
 \label{eq:RSB-operations-2}
 \{e^{i\tfrac{2\pi}{d_2 d_1 N}\hat{n}}, e^{i\tfrac{\pi}{d_2 d_1}\left(\frac{\hat{n}^2}{{d_1} N^{2}}-\beta\frac{\hat{n}}{N}\right)},e^{i\tfrac{2\pi}{d_2 {d_1}^2 N^2}\hat{n}_{k}\hat{n}_{l}},\text{PM}\}.
\end{align}
The parameter $\beta$ takes values $\beta=0$ for even $d_2$, and $\beta=1$ for odd $d_2$, for with $d_2$ a natural number.

To immediately grasp the implications of these results, let us focus on GKP circuits and the useful case of $d_1 =2$. In this case, by embedding the qubit codewords into a qudit of dimension $d_2=2a^2$ (with $a \in \mathbb{N}$), we will show that it is possible to efficiently simulate circuits generated by the set of operations given in Eq.~(\ref{eq:GKP-operations}), with $\alpha = \sqrt{\pi}/a$. That is, circuits that include momentum and position displacements in phase space by arbitrary fractions of $\sqrt\pi$ can be simulated. This recognizes as classically efficiently simulatable a variety of operations beyond the qubit-encoded stabilizer ones, as operations in the qubit logical space correspond only to displacements of integer multiples of $\sqrt\pi$.
Note that $a$ can be made arbitrarily large, yielding simulatable displacements that are arbitrarily small.
As already mentioned, these circuits are efficiently simulatable even though theorems for the simulatability of CV and DV systems cannot be immediately applied, given that the input states possess negative Wigner functions and the encoded Clifford group operations for $d_2$-dimensional systems take the logical qubit states outside of their codespace.

For RSB codes, the simulatable operations are listed in Eq.~(\ref{eq:RSB-operations-1}) and Eq.~(\ref{eq:RSB-operations-2}).
We start by considering the extension of the definition of order-$N$ bosonic rotation codes that was given for qubits~\cite{grimsmo2020} to include higher-dimensional qudits, and define the corresponding logical Clifford operations, along the lines of
Refs.~\cite{vourdas2004,raynal2010,albert2016,bergmann2016a,li2017,albert2018,albert2019,albert2019a}.
The resulting circuits (see Fig.~\ref{fig:circuitrsb}) are efficiently simulatable even though all their components (input states, gates, and measurements) display negativities in their respective Wigner functions. The qubit RSB codewords may be encoded in a $d_2$-dimensional codespace of a qudit in two ways. The first alternative is similar to the GKP case, and corresponds to choosing $d_2=2a^2$ (for $a \in \mathbb{N}$), together with the additional constraint $M=N/a$, with $N$ and $M$ the rotation symmetry of the qubit and qudit, respectively.
The second option consists in the encoding of an $N$-fold symmetric computational basis $0$-state of a qubit, denoted as $\ket{\logic{0_2}_{;N,\varphi}}$,  into a
$M$-fold symmetric qudit state $|\logic{+_{d_2}}_{;M,\varphi}\rangle=\sum_{j=0}^{d_2-1}|\logic{j_{d_2}}_{;M,\varphi}\rangle/\sqrt{d_2}$ of any dimension $d_2$, with the additional condition $M=2N$. In both cases, the families of circuits that can be efficiently simulated comprises a much larger variety of operations than the qubit stabilizer circuits. We will describe them in detail later on and they are summarized in Table~\ref{tab:RSBoperations}.

The main difference between GKP and RSB circuits is that while in the former Wigner function negativities are only present in the encoded input codewords, in the latter case some of the encoded Clifford operations are also highly non-Gaussian and Wigner negative. This is due to the fact that for RSB codes it is no longer true that Clifford operations map onto Gaussian operations, in contrast to GKP codes.

\section{Discrete-variable systems: Qubits and qudits}
\label{sec:qubits-qudits}
In this Section, we review the standard notion of qubits and qudits, as well as the corresponding definitions of the Pauli group and Clifford group, along with the notion of stabilizer states. We also recall the encoding of qubits into higher-dimensional systems.

\subsection{Pauli group and Clifford group for qubits}

Throughout this paper, we will refer to qubits as the traditional two-level quantum system, which can be written as $\ket\psi=\alpha\ket 0 +\beta \ket 1$ for some $\alpha,\beta \in \mathbb C$, $\abs{\alpha}^2+\abs{\beta}^2=1$, and $\ket 0$ and $\ket 1$ define the computational basis.
The Pauli group for a single qubit can be defined as $\mathcal P_2~=~\{\pm i^{u}X^{v}Z^{w}~:~u,v,w\in\mathbb{Z}_2\}$ where the Pauli operators are given by $X=\ket 0 \bra 1+\ket 1\bra 0$ and $Z=\ket 0 \bra 0-\ket 1\bra 1$. It can also be expressed in terms of the group's generators (\textit{i.e.}, the operators which  generate the entire group by successive application) as $\mathcal P_2~=~\langle i I_2,X,Z\rangle$, where $I_2$ denotes the $2$-dimensional identity operator.
The Pauli group for $n$ qubits is given by $\mathcal P_2^{n}=\bigotimes^{n}_{j=1} \mathcal P_2$~\cite{nielsen2000}.

Formally, the Clifford group for a set of $n$ qubits is defined as
\begin{align}
 \mathcal C_2^{n} = \{ Q : QUQ^\dagger \in \mathcal P_2^{n} \quad \forall \quad U \in \mathcal P_2^{n} \} .
\end{align}
It is well known that the Clifford group for qubits can be generated by $\mathcal C_2^n=\langle H,S,\textsc{cnot} \rangle$, where $H$ is the Hadamard gate, $S$ is the phase gate, and $\textsc{cnot}$ is the two qubit controlled-$\textsc{not}$ gate~\cite{aaronson2004}. This notation implicitly assumes that $H, S, \textsc{cnot}$ act on any qubit or pair of qubits.

\subsection{Pauli group and Clifford group for qudits}
\label{subsec:Clifford_qudit}
Systems of higher (finite) dimensions than qubits are similarly defined. A $d$-dimensional qudit is defined as $\ket{\psi}=\sum_{j=0}^{d-1}\alpha_j\ket j$ for $\alpha_j\in \mathbb C$, $\sum_{j=0}^{d-1} |\alpha_j|^2=1$, and $\ket j$ being the computational basis states.
The imaginary phase in higher dimensions takes the form of a primitive $d$-th root of unity,
\begin{align}
 \label{eq:omega_d}
 \omega_d=e^{2\pi i/d} .
\end{align}
The $d$-dimensional Pauli group on one qudit is defined as $\mathcal{P}_{d}=\{\omega_{D}^{u}X_{d}^{v}Z_{d}^{w} : v,w\in \mathbb Z_{d},  u\in\mathbb{Z}_{D}\}$ with
\begin{align}
 \label{eq:dandD}
 D=
 \begin{cases}
  d,  & \text{for odd } d,  \\
  2d, & \text{for even } d,
 \end{cases}
\end{align}
where the $d$-dimensional Pauli generators are~\cite{gottesman1999a,farinholt2014,gheorghiu2014}
\begin{align}
 \label{eq:pauli-dim-d}
  & X_d=\sum_{j=0}^{d-1} \ket{j+1\bmod d}\bra{j} ,  \nonumber \\
  & Z_d=\sum_{j=0}^{d-1} \omega_d^j \ket j \bra j .
\end{align}
The operators $X_d$ and $Z_d$ have order $d$, meaning that $X_d^d=I_d$ and $Z_d^d=I_d$ both equal the $d$-dimensional identity.
As in the qubit case, the group can be expressed in terms of the generators, $\mathcal P_d=\langle\omega_{D} I_d, X_d, Z_d \rangle$~\cite{farinholt2014}. The Pauli group is an essential component of the stabilizer formalism, which can be used to systematically correct errors in quantum computation~\cite{gottesman1996,gottesman1997,poulin2005,nielsen2000}. The Pauli group for $n$ qudits is given by $\mathcal P_d^{n}=\bigotimes^{n}_{j=1} \mathcal P_d$.

Correspondingly, the Clifford group is generalized as
\begin{align}
 \label{eq:Clifford-d}
 \mathcal C_d^{n} = \{ Q : QUQ^\dagger \in \mathcal P_d^{n} \quad \forall \quad U \in \mathcal P_d^{n} \}.
\end{align}
Analogously to the qubit case, the Clifford group over $n$, $d$-dimensional qudits can be understood as the group generated by
$\mathcal C^{n}_d=\langle F_d,S_d,\textsc{sum}_d \rangle$~\cite{farinholt2014}, where we define the qudit gates as
\begin{align}
 F_d = \frac{1}{\sqrt d}\sum_{j,k=0}^{d-1} \omega_d^{jk}\ket{k}\bra{j} ,
\end{align}
\begin{align}
 \label{eq:phase-gate-clifford}
 S_d = \sum_{j=0}^{d-1} \omega_d^{j^2/2}\eta_d^{-j}\ket{j}\bra{j} ,
\end{align}
and,
\begin{align}
 \textsc{sum}^{(k,l)}_d = \sum_{i,j=0}^{d-1} \ket{i}^{(k)} \prescript{(k)}{}{\bra{i}} \otimes \ket{i+j}^{(l)} \prescript{(l)}{}{\bra{j}},
\end{align}
whereby $\omega_d$ is defined in Eq.~(\ref{eq:omega_d}),  $\eta_d=\omega_{D}\omega_{2d}^{-1}$ and addition is defined as modulo $d$.
Note that for $d=2$ we have $F_2=H$, $S_2=S$ and $\textsc{sum}_2=\textsc{cnot}$.

\subsection{Stabilizer formalism and encoding lower-dimensional qudits into higher-dimensional qudits}
\label{subsec:stabiliser-formalism}

The stabilizer formalism can be used to simulate the evolution of stabilizer states under the action of Clifford group operations. The key to the ability to simulate these states efficiently is that each stabilizer state can be represented as a corresponding group named the stabilizer group.
A stabilizer state is a simultaneous eigenvector of eigenvalue 1, for all elements in a stabilizer group consisting of $d^n$ commuting elements of the Pauli group $\mathcal P_d^n$ which contains exactly one identity operator~\cite{hostens2005}. This group can be specified in terms of its group generators. Each stabilizer state is represented by a unique stabilizer group which transforms under Clifford operations such that the evolution of the stabilizer state can be tracked using the stabilizer group generators~\cite{hostens2005}.

Similarly, it is possible to define a stabilizer group associated to a subspace and, in particular, one can define stabilizer generators for the logical space or codespace of qubits and qudits, in either DV or CV codes. The codespace is as well spanned by the codewords, which are logical encoded states in that subspace.

In the traditional stabilizer formalism, one encodes the logical qubit information across $n$ physical qubits,  defining a code. An alternative to this traditional formalism has been introduced in Ref.~\cite{gottesman2001}, where a single logical qubit is instead encoded in a single higher-dimensional physical qudit. A continuous bosonic code is then obtained as an extension of this encoding strategy to infinite-dimensional systems. Here we  review the mapping of qubits into qudits, as it will be relevant for later results of Sec.~\ref{sec:gkp-qunits-to-qudits}.

To redundantly encode the logical qubit information inside a physical qudit, the qubit state can be encoded evenly across the span of the qudit encoding space. The logical qubit basis states become
\begin{align}
 \label{eq:qubit-qudit-DV}
 \ket{\logic{j_2}}=\frac{1}{\sqrt{a_2}}\sum_{k=0}^{a_2-1}\ket{(2k+j)a_1}_d
\end{align}
for some integer $a_1$ and $a_2$ such that $d=2 a_1 a_2$, and where $j$ is chosen from $j=0,1$.
Here, and throughout our manuscript, we use bold fonts to represent logical encoded states and operations, with the bold subscript inside the ket denoting the logical space dimension.
We add an additional subscript outside the ket in the right-hand side to indicate the dimension of the physical qudit, in contrast with the bold subscript.

For example, if we consider $d=8$ and $a_1=a_2=2$, we can encode the logical qubit state $\ket{\mathbf{0_2}}$  in dimension 8 as
\begin{align}
 \label{eq:qubit-qudit-DV-example}
 \ket{\mathbf{0_2}}=\frac{1}{\sqrt{2}}(\ket{0}_8+\ket{4}_8) .
\end{align}

The above scheme can be extended to encode a $d_1$-dimensional qudit within a larger $d_2$-dimensional qudit
\begin{align}
 \label{eq:qunit_qudit_DV}
 \ket{\logic{j_{d_1}}}=\frac{1}{\sqrt{a_2}}\sum_{k=0}^{a_2-1}\ket{(k {d_1}+j)a_1}_{d_2},
\end{align}
where $j=0,\dots,d_1-1$. In this case, the encoded Pauli operators are the $d_2$-dimensional operators applied $a_1$ and $a_2$ times respectively,
\begin{align}
  & \logic{X_{d_1}}=X_{d_2}^{a_1}, \nonumber \\
  & \logic{Z_{d_1}}=Z_{d_2}^{a_2} .
\end{align}

The $d_1$-dimensional qudit information is therefore redundantly encoded in a minimum of $d_2=d_1a_1a_2$. 

For symmetric encodings, \textit{i.e.}, $a_1 = a_2$, Clifford operations in $d_2$-dimensions also act as their equivalent Clifford operators in $d_1$-dimensions ~\cite{farinholt2014}. For an example, see Appendix~\ref{app:stabilizer_preservation}.

It is important to observe that the qudit state on the right-hand side of Eq.~(\ref{eq:qunit_qudit_DV}) is always a stabilizer state in dimension $d_2$, \textit{i.e.}, with respect to the Clifford group operations in $d_2$.
Indeed, it is possible to define a circuit which generates it using two qudits initialized in $\ket{0}_{d_2}$~\cite{beaudrap2013} (see Appendix~\ref{app:generating_encoded_qubit} for an explicit construction in the case of $d_1 = 2$).
Equivalently, we can observe that the stabilizer generators of the logical subspace for the $d_1$-dimensional system are defined as $X_{d_2}^{d_1a_1}$ and $Z_{d_2}^{d_1a_2}$, and that the stabilizer group which specifies the state in Eq.~(\ref{eq:qunit_qudit_DV}) is given as $\mathcal S = \langle X_{d_2}^{d_1a_1},Z_{d_2}^{d_1a_2} \rangle$.
The state is an eigenstate of all of the elements generated by this group with eigenvalue $1$, and therefore is a stabilizer state in dimension $d_2$.

In the previous case, we always consider a composite dimension $d_2$. The analysis differs for prime $d$ dimensions, where it can be shown that a target single qudit state is a stabilizer state if and only if there exists a single qudit Clifford group operation which takes the $\ket 0_{d}$ state to the target state~\cite{gottesman1999a}. In prime dimensions --- including the simplest and most common $d=2$ case --- the stabilizer group can always be generated by a single Pauli group operator~\cite{gheorghiu2014}.

\section{Efficient simulation of discrete-variable systems}
\label{sec:DV-theorems}
Among the extensive literature related to efficient classical simulation of non-trivial classes of quantum circuits in DVs~\cite{vidal2003,aaronson2004,van-den-nest2010,ohliger2012,mari2012,veitch2012,pashayan2015,bravyi2016,bennink2017, bu2019efficient}, we will focus on the class of stabilizer circuits.

In the literature, we can distinguish between two notions of classical simulation which are referred to as weak and strong simulation. Weak simulation refers to the process of using a classical computer to sample the output distribution with high accuracy, whereas strong simulation refers to the computation of the probability distribution of the output measurement for any possible measurement outcome~\cite{van-den-nest2010,jozsa2014, koh2017further}.

Stabilizer circuits are quantum circuits composed of Clifford operations and Pauli measurements. They arise in several applications as quantum error-correcting codes, and their evolution is known to be classically efficiently simulatable when applied to initial computational basis states~\cite{gottesman1999}.

In this Section, we briefly review the known results on the efficient simulation of codes consisting of $d$-level systems. We start by recalling the seminal Gottesman-Knill theorem for the strong classical simulation of a certain class of quantum circuits applied onto $2$-level qubit systems. Later, we review extended theorems for higher-dimensional systems of arbitrary integer dimension $d$, which will later be used in order to derive our results.

\subsection{Gottesman-Knill theorem}
\label{sec:DV-theorems-GK}
The Gottesman-Knill theorem outlines a subset of quantum computations and measurements that can be efficiently simulated on a classical computer. In essence, by limiting the range of possible computations to  Clifford operations, it is possible to efficiently track the result of these operations and to predict the probability distribution of the output of Pauli measurement over all qubits~\cite{gottesman1999,aaronson2004,jozsa2014}.

The Gottesman-Knill theorem utilizes the stabilizer formalism to track the information of a quantum state by in turn storing the information of its stabilizer group. The initialized quantum register for $n$ qubits $\ket{000\dots 0}$ is a stabilizer state admitting a stabilizer group specified by the generator $Z_j$ for each qubit $j$. There are exactly $n$ stabilizer generators for a system of $n$ qubits.

Clifford operations can be taken into account by updating the $n$ stabilizer group generators.
By tracking how the stabilizer generators transform under Clifford operations, the final stabilizer group can be reproduced. Simulating measurements involves identifying whether the measured operator commutes or anticommutes with each element of the stabilizer group. This information allows for the construction of the probability density of the measurement of any Pauli operator~\cite{nielsen2000}. The complexity of simulating the action of Clifford operations on qubits initialized in the Pauli basis, and with a single or constant number of measurements is known to lie within the complexity class P, meaning that it can be calculated in $poly(n)$ time~\cite{aaronson2004}.

\subsection{Theorems for higher-dimensional systems}
\label{sec:DV-theorems-higher}

While it was originally developed for qubits, this formalism can be easily extended to prime dimensions,  and the Clifford operations defined in Sec.~\ref{subsec:Clifford_qudit}) are also efficiently simulatable in these higher dimensions~\cite{gottesman1999a}.
Any prime $d$-dimensional system consisting of $n$ qudits has $n$ stabilizer generators.

Note that, in the case of prime dimensions, it is possible to utilize the non-negativity of the discrete Wigner function~\cite{gross2006} to identify stabilizer states. This yield to alternative theorems for the simulatability of Clifford circuits~\cite{veitch2012,veitch2014}, and even other algorithms based on low-rank stabilizer decompositions cover the situation of quantum circuits with few non-Clifford gates for qubits and odd-dimensional systems~\cite{dam2011,bravyi2016a,bravyi2016,bravyi2019,huang2019}. The runtime of these methods scales exponentially with the number of non-Clifford gates. Further results demonstrate  the simulatability of Clifford circuits with input mixed non-stabilizer states, where the running time increases with the purity of the input state~\cite{bu2019efficient}.

When we relax the condition of $d$ prime, we find that many of these nice properties do not hold, including the ability to write the stabilizer states in a general form as non-negatively Wigner represented states~\cite{appleby2008,dam2011}.
Therefore, we must rely on more general theorems to identify whether a qudit system is simulatable~\cite{hostens2005,beaudrap2013,bermejo-vega2014}.

For systems of arbitrary dimensions, Refs.~\cite{hostens2005,farinholt2014} outline a method to track the evolution of qudit stabilizer states under the operation of Clifford group operators. 
By starting from the $\ket{000\dots 0}$ state of $n$ qudits, any other stabilizer state can be reached using Clifford operations and measurements. It is possible to track Clifford group operations using the stabilizer formalism. Rather than using an exponentially large matrix, the stabilizer formalism uses a $2n\times2n$ matrix with entries within $\mathbb Z_d$ and a $2n$-dimensional vector with entries in $\mathbb Z_{2d}$, which details the Clifford transformations required to reach the described state.

As already mentioned, the Clifford operations can be generated using two single qudit gates and one two-qudit gate: the Fourier transform $F_d$, the phase gate $S_d$ and the SUM gate $\textsc{sum}$~\cite{farinholt2014}.
In higher-dimensional systems, Pauli operators are non-Hermitian, and therefore, by Pauli measurements we refer to Hermitian projection operators that collapse the state of the system onto the corresponding eigenspaces of the Pauli operators with certain eigenvalues~\cite{gottesman1999a,beaudrap2013}.

Strong simulation of circuits consisting of measurement of the final state of all qudits after $d$-dimensional Clifford operations are applied to an initial stabilizer state in arbitrary dimension $d$ is contained within $\text P$~\footnote{
 Simulating stabilizer circuits with a single measurement for qubits is complete for the complexity class $\oplus\textsc{L}$, which are the class of problems equivalent to solving a linear system over $\mathbb Z_2$~\cite{damm1990}. It is also known that $\oplus\textsc{L}\subseteq \textsc{P}$~\cite{aaronson2004}, and hence simulating stabilizer circuits with a constant number of measurements is contained within $\textsc{P}$.
 Simulating stabilizer circuits with a constant number of measurements in arbitrary $d$-dimensional qudit systems is complete for the complexity class $\textsc{coMod}_d\textsc{L}\subseteq \textsc {P}$~\cite{beaudrap2013}. Ref.~\cite{van-den-nest2013} shows that strong simulation of circuits consisting of measurement of the final state of all qudits is contained within $\text P$.} meaning that the evolution and measurement of all qudits can be simulated in $poly(n)$ time.

In the following, to derive our results for CV architectures, we are going to exploit in particular the embedding of logical $d_1$-dimensional qudits into $d_2$-dimensional qudits with both composite and prime dimensions. Firstly, for both the GKP code and RSB codes, we can use the encoding of a qubit in a qudit of dimension $d=2a^2$, therefore with $d$ composite. We will be able to exploit results for prime dimension only within a particular encoding method in RSB codes, that we outline in Sec.~\ref{sec:qubit-into-qudit-RBC}.

\section{Bosonic codes: Qubits and qudits in continuous-variable systems}
\label{sec:bosonic}
It is also possible to encode qubits and qudits in CV systems to protect the quantum information in the finite-dimensional encoded states. Several such schemes have been extensively investigated~\cite{gottesman2001, vourdas2004, raynal2010, mirrahimi2014, albert2016, bergmann2016, michael2016, grimsmo2020}. We focus in particular on GKP and RSB codes.

\subsection{Discrete translation-symmetric codes}
Among these schemes designed to be robust against unwanted errors on a logical qubit, we focus in this Section on the translation-symmetric GKP code. This code maps the state of a logical DV system across a grid in the phase space of an oscillator. This mapping protects the quantum information against shifts in both position and momentum~\cite{gottesman2001,menicucci2014}, and has been successfully demonstrated experimentally~\cite{fluhmann2018,campagne-ibarcq2019,fluhmann2019}.
In the following discussion, we will focus on the encoding whereby the DV-system (\textit{e.g.}, a qubit) state is encoded across a square lattice. Notice that it is possible to achieve an even larger number of correctable displacements by encoding the states across a hexagonal lattice~\cite{gottesman2001}, however, for the sake of simplicity, we will use the square lattice. Our results can be easily extended to other  lattices.

\subsubsection{GKP code: Codewords and operations for qubits}
\label{sec:gkp-qubits}
In the ideal case of infinite squeezing, we can encode a logical qubit --- as said, a DV system of dimension $d=2$ --- into a GKP state as $\ket{\logic{\psi_2}}=\alpha\ket{\logic{0_2}}+\beta\ket{\logic{1_2}}$, for arbitrary complex amplitudes $\alpha$ and $\beta$, whereby the computational basis codewords are defined as
\begin{align}
 \label{eq:code-words-qubit}
  & \ket{\logic{0_2}} = \sum_{s\in\mathbb Z} \ket{2s\sqrt{\pi}}_{\hat q}, \nonumber \\
  & \ket{\logic{1_2}} = \sum_{s\in\mathbb Z} \ket{(2s+1)\sqrt{\pi}}_{\hat q} ,
\end{align}
where $\ket{\cdot}_{\hat q}$ denote non-normalizable eigenstates of the position operator $\hat q$. The use of this subscript outside the ket refers to the physical dimension of the system, analogous to the notation introduced in Eq.~(\ref{eq:qubit-qudit-DV}). The Pauli logical operations for a qubit are then given by
\begin{align}
 \label{eq:clifford-gkp0}
  & \logic{Z_2} = e^{i{\hat q}\sqrt{\pi}}, \nonumber \\
  & \logic{X_2} = e^{-i{\hat p}\sqrt{\pi}}.
\end{align}

A notable aspect of the GKP encoding is that Clifford operations on the logical encoded states are described by  Gaussian operations for the underlying bosonic modes, and thus they are operations characterized by a non-negative Wigner function.
In particular, of relevance for our purposes are the following Clifford operations: $\logic{Z_2}$, $\logic{X_2}$, Hadamard, phase gate, and $\textsc{\logict{cnot}}$. Their GKP-encoded counterparts are given respectively by Eq.~(\ref{eq:clifford-gkp0}), by the Fourier transform,
\begin{align}
 \label{eq:clifford-gkp2}
 \logic{F}= e^{i\frac \pi 4 \left( \hat p^2 + \hat q^2\right)} ,
\end{align}
by the GKP-encoded phase gate (see Appendix~\ref{app:sum-gate})
\begin{align}
 \label{eq:clifford-gkp3}
 \logic{S}= e^{i \frac{\hat q^2}{2}} ,
\end{align}
and by the GKP-encoded \textsc{sum} gate,
\begin{align}
 \label{eq:clifford-gkp4}
 \textsc{\logict{sum}}^{(k,l)}= e^{-i\hat{q}_k \hat{p}_l} .
\end{align}

For completeness, although not included in the set of operations that we will address when deriving our simulatability results, we mention that in order to achieve universal quantum computation on the encoded qubits one might add to the set of Clifford operations presented above the non-Clifford gate $T$, defined along with its GKP-encoded counterpart $\logic{T}$ in Appendix~\ref{app:t_gates}.

\subsubsection{GKP code: Codewords and operations for qudits}
We can similarly use the GKP mapping to encode a $d$-dimensional qudit on CV systems such that the computational basis codewords are~\cite{gottesman2001}
\begin{align}
 \label{eq:code-words-2}
 \ket{ \logic{j_d}} = \sum_{s\in\mathbb Z} \ket{\sqrt{\tfrac{2\pi}{d}}(j+ds)}_{\hat q} .
\end{align}
The logical operations acting on the qudit are, in this case, given by
\begin{align}
 \label{eq:clifford-gkp1}
  & \logic{Z_d} = e^{i{\hat q}\sqrt{\frac{2\pi}{d}}}, \nonumber \\
  & \logic{X_d} = e^{-i{\hat p}\sqrt{\frac{2\pi}{d}}}.
\end{align}

In arbitrary dimension as well, Clifford operations on the logical encoded states are described by Gaussian operations, and are therefore positive Wigner operations. The $d$-dimensional Clifford GKP-encoded group include the set of finite displacements in the $\hat q$ and $\hat p$ quadrature given by $\logic{Z_d}$ and $\logic{X_d}$ in Eq.~(\ref{eq:clifford-gkp1}) respectively (to different dimension correspond different amplitude of the displacements), the Fourier transform Eq.~(\ref{eq:clifford-gkp2}), the phase gate
\begin{align}
 \label{eq:clifford-gkp3-bis}
 \logic{S}= e^{\frac{i}{2} \left(\hat q^2 - 2c \hat q \right)} ,
\end{align}
whereby $c=0$ for even $d$-dimensions and $c=\sqrt{\pi/2d}$ for odd $d$-dimensions, and the \textsc{sum} gate is given in Eq.~(\ref{eq:clifford-gkp4}). In Appendix~\ref{app:sum-gate}, we provide an explicit derivation of Eq.~(\ref{eq:clifford-gkp3-bis}), given the definition of the phase gate in Eq.~(\ref{eq:phase-gate-clifford}).

Both Eq.~(\ref{eq:code-words-qubit}) and Eq.~(\ref{eq:code-words-2}) represent an ideal encoding in non-normalizable states. Their wave and Wigner function are characterized by infinite series of Dirac-delta peaks. On the other hand, realistic finite-squeezing GKP states are given by a series of normalized Gaussians with width $\Delta$, enveloped by a Gaussian of width $\delta^{-1}$. The realistic wave-functions are therefore given by (in general dimension $d$, including the qubit case $d = 2$)
\begin{align}
 \bra{q}\ket{\logic{j_d}} \propto \sum_{s\in\mathbb Z}  e^{-(j+ds)^2\pi\delta^2/2}e^{-\left(q-(j+ds)\sqrt{\pi}\right)^2/2\Delta^2}
\end{align}
up to some normalization dependent on each logical state.

In both cases of ideal and realistic encoding, the states are non-Gaussian and Wigner negative. The negativity of the Wigner function of encoded GKP states, both stabilizers and non-stabilizer, has been studied in Refs.~\cite{garcia-alvarez2019,yamasaki2019}. For the sake of simplicity, we use the ideal encoding in our derivations.

\subsection{Discrete rotation-symmetric codes}
\label{sec:rot_codes}
A large group of single-mode bosonic error-correcting codes, including cat and binomial codes, can be classified as rotation-symmetric bosonic codes~\cite{grimsmo2020}. The stabilizer generator of these RSB codes is given by a discrete rotation symmetry, that is, the codespace is a $+1$ eigenspace of the $N$-fold rotation symmetry operator
\begin{align}
 \hat{R}_N = e^{i \frac{2\pi}{N} \hat{n}}.
\end{align}
Here we introduce an extension of this encoding, as well as of the corresponding logical operations,   to the case of RSB-encoded qudits in arbitrary dimension, of which RSB-encoded qubits~\cite{grimsmo2020} are a particular case.

\subsubsection{RSB code: Codewords and operations for qubits}

For RSB-encoded qubits, the logical operation $Z_2$ is defined as
\begin{align}
 \logic{Z_2} = \hat{R}_{2N} = e^{i \frac{\pi}{N} \hat{n}}.
\end{align}
It satisfies the relation $\logic{Z_2} |\logic{j_2}_{;N,\varphi}\rangle = (-1)^{j} |\logic{j_2}_{;N,\varphi}\rangle$ when applied to codewords of the form
\begin{align}
 |\logic{j_2}_{;N,\varphi}\rangle = \frac{1}{\sqrt{{\mathcal{N}}_j} } \sum_{m=0}^{2N-1} (-1)^{jm} e^{i \frac{m \pi}{N} \hat{n}} |\varphi\rangle,
\end{align}
where $j=0,1$, ${\mathcal{N}}_j$ are normalization constants, and $|\varphi\rangle$ is a primitive state characteristic of the particular encoding (\textit{e.g.}, a coherent state for the case of cat codes).

It is possible to define the Clifford generators (\textit{i.e.}, Hadamard, phase, and controlled-$Z$ gates) in RSB codes for qubits~\cite{grimsmo2020}. We do not list them here though, since in  Sec.~\ref{sec:qudit_RSB} we define the RSB Clifford operations for the general case of qudits, of which qubits are a particular case.

For universal quantum computation, one could add the non-Clifford gate $T$ defined in Appendix~\ref{app:t_gates}. As it occurs with GKP code, this operation cannot be simulated efficiently classically with our method.

\subsubsection{RSB: Codewords and operations for qudits}
\label{sec:qudit_RSB}
Here we consider the extension of the definition of order-$M$ bosonic rotation codes and operations for qubits that was developed in Ref.~\cite{grimsmo2020} to include higher-dimensional systems or qudits.
Our definition of a RSB qudit is analogous to previous definitions in the literature~\cite{vourdas2004, raynal2010, albert2016, bergmann2016a, li2017, albert2018, albert2019, albert2019a}.
For example, the $X$ operator may correspond to the discrete position operator of a particle on a ring~\cite{vourdas2004,albert2016}.

The codespace, as in the qubit case, is given by a $+1$ eigenstate of a rotation operator $\hat{R}_M$ with $M$-fold rotation symmetry. We define now the generalized Pauli operator acting on a $d$-dimensional system as
\begin{align}
 \label{eq:zd_rot}
 \logic{Z_d}= \hat{R}_{dM}=e^{i\tfrac{2\pi}{dM}\hat{n}}.
\end{align}
The computational basis states for a qudit satisfy the relation $\logic{Z_d}|\logic{j_d}\rangle = \omega_d^{j}|\logic{j_d}\rangle$ with $\omega_d$ as defined in Eq.~(\ref{eq:omega_d}). Therefore, for the encoding in dimension $d$, they are described by
\begin{align}
 \label{eq:codewordrbc}
 |\logic{j_d}_{;M,\varphi}\rangle = \frac{1}{\sqrt{{\mathcal{N}}_j}} \sum_{m=0}^{dM-1}
 \omega_d^{-jm} e^{i\frac{2\pi}{dM}m\hat{n}}|\varphi\rangle,
\end{align}
with ${\mathcal{N}}_j$ the corresponding normalization factor. The Fock-space structure of the qudit codewords shows a spacing $dM$ (\textit{i.e.}, it depends on the dimension $d$) and  rotation symmetry $M$, as shown in Appendix~\ref{app:fock_qudit}.

Alternatively, it is possible to describe the qudit states in terms of the $\logic{X_{d}}$ basis states $|\logic{u^0_d}_{;M,\varphi}\rangle$. The relation between $\logic{Z_{d}}$ basis and the $\logic{X_{d}}$ basis for a qudit of dimension $d$ is
\begin{align}
 \label{eq:Xd_codewords}
 |\logic{u^k_d}\rangle = \frac{1}{\sqrt{d}} \sum_{j=0}^{d-1} \omega_d^{-kj} |\logic{j_{d}}\rangle,
\end{align}
with $j\in\mathbb{Z}_d$, and $k\in\mathbb{Z}_d$, such that, $\logic{X_{d}}|\logic{u^k_d}\rangle = \omega_d^{k} |\logic{u^k_d}\rangle$, and $\logic{Z_{d}}|\logic{j_d}\rangle = \omega_d^{j} |\logic{j_d}\rangle$.
For the sake of clarity, we have omitted the primitive function $\varphi$ dependence, and the $M$ label related to the fold rotation symmetry.

It is possible to define a universal set of operations for the encoded qudits. In particular, for our results we focus on a set of gates able to generate the Clifford group, namely, $\langle \logic{S_{d}},\logic{F_{d}},\logic{C_{Z}}\rangle$, and the generalized Pauli gate $\logic{Z_{d}}$.

As it will become clear in the forthcoming Sections, the qudit dimension $d_2$ that we will consider may be either even, if for example we choose to embed a qubit (\textit{i.e.}, $d_{1} = 2$) in a higher dimension $d_2$, such that $d_{2}=d_{1}a^2$, or odd, if we choose to encode another $d_1$-dimensional system.
Moreover, a second alternative method allows us to interpret $d_1$-dimensional states in any dimension $d_2$. Then, we need to consider the phase gate $\logic{S_{d}}$ for any dimension $d$ even or odd.
The phase gate for qudits is defined in Eq.~(\ref{eq:phase-gate-clifford}), which in RSB codes corresponds to the self-Kerr interaction
\begin{align}
 \label{eq:phase_rot_qudit}
 \logic{S_{d}} = e^{i\tfrac{\pi}{d}\left(\frac{\hat{n}^2}{M^{2}}-\beta\frac{\hat{n}}{M}\right)},
\end{align}
with $\beta=0$ for even dimensions, and $\beta=1$ in the case of odd dimensions (see derivation in Appendix~\ref{app:rot_gates_details}).

The entangling controlled-$Z$ gate, $\logic{C_{Z}}$, for qudits of dimension $d$ is defined on the computational basis states $j$ and $l$ as
\begin{align}
 \logic{C_{Z}}|\logic{j_d}\rangle |\logic{\ell_d}\rangle = e^{i\frac{2\pi}{d}j \ell} |\logic{j_d}\rangle |\logic{\ell_d}\rangle.
\end{align}
Therefore the corresponding RSB-encoded operation $ \logic{C_{Z}}^{({k}_{N},{s}_{M})}$ is a controlled rotation between two modes $k$ and $s$ with $N-$ and $M-$fold rotation symmetry, respectively,
\begin{align}
 \label{eq:controlZ_rot_qudit}
 \logic{C_{Z}}^{({k}_{N},{s}_{M})} = e^{i\tfrac{2\pi}{dNM}\hat{n}_{k}\hat{n}_{s}}.
\end{align}

The $\logic{F_{d}}$ gate can be defined by its action on an arbitrary RSB-encoded qudit $|\logic{\psi_d}\rangle=\sum_{k=0}^{d-1} \alpha_k |\logic{k_d}\rangle$,
\begin{align}
 \label{eq:fourier_rbc}
 \logic{F_{d}} \ket{\logic{\psi_d}} = \frac{1}{\sqrt{d}}\sum_{j,k=0}^{d-1} e^{i\tfrac{2\pi}{d}kj} \alpha_k |\logic{j_d}\rangle.
\end{align}

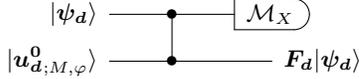
\begin{figure}[htpb]
 \centering
 $$
  \Qcircuit @C=2.5em @R=1.2em {
  \lstick{|\logic{\psi_d}\rangle}  &  \ctrl{1}  & \measureD{\mathcal{M}_{X}}\\
  \lstick{|\logic{u^0_d}_{;M,\varphi}\rangle} & \control \qw & \rstick{\logic{F_{d}}|\logic{\psi_d}\rangle} \qw
  }
 $$
 \caption{Circuit structure for implementing the $\logic{F_{d}}$ gate on an arbitrary qudit state $|\logic{\psi_d}\rangle$.
  It consists of two qudits initialized in $|\logic{\psi_d}\rangle$ and $|\logic{u^0_d}_{;M,\varphi}\rangle \equiv |\logic{+_d}_{;M,\varphi}\rangle$, respectively, followed by a $\logic{C_{Z}}$ gate, and a phase measurement $\mathcal{M}_{X}$ on the first qudit~\cite{helstrom1969,holevo2011,grimsmo2020}.
  The measurement is a projection operation on the $\logic{X_{d}}$ eigenstates, which can be obtained with $1/d$ probability each. Upon  post-selection on the outcome $+1$, we obtain an effective Fourier gate as given in Eq.~(\ref{eq:fourier_rbc}).
 }
 \label{fig:teleportedFourier}
\end{figure}

In order to implement physically this gate, we follow the circuit structure in Fig.~\ref{fig:teleportedFourier}, which represents a $d$-dimensional standard gate-teleportation gadget~\cite{nielsen1997programmable,gottesman1999b,nielsen2000}.
We consider an auxiliary qudit initialized in the $\logic{X_d}$ $+1$ eigenstate $|\logic{u^0_d}_{;M,\varphi}\rangle$, thus we start from
\begin{align}
 |\logic{\Psi}_{in}\rangle=\frac{1}{\sqrt{d}}\sum_{k=0}^{d-1} \alpha_k |\logic{k_d}\rangle \sum_{j=0}^{d-1} |\logic{j_d}\rangle ,
\end{align}
and we apply initially a $\logic{C_{Z}}$ gate before the projective measurement in the $\logic{X_{d}}$ basis,
\begin{align}
 |\logic{\Psi}_f\rangle=\logic{C_{Z}}|\logic{\Psi}_{in}\rangle = \frac{1}{\sqrt{d}}\sum_{j,k=0}^{d-1} \alpha_k e^{i\tfrac{2\pi}{d}kj} |\logic{k_d}\rangle |\logic{j_d}\rangle .
\end{align}

Conditioned on getting the outcome corresponding to the state $|\logic{u^0_d}_{;M,\varphi}\rangle$ when measuring the first qubit, we find the second qubit in the state given in Eq.~(\ref{eq:fourier_rbc}).
Post-selection is needed, and therefore, the probability of the conditioning result must be non zero. In this case, the conditional probability describing the post-selection is $1/d$ for any $\logic{X_{d}}$ eigenstate $|\logic{u^n_d}_{;M,\varphi}\rangle$, with $n=0,1,\dots, d-1$. It does not depend on the input state $|\logic{\psi_d}\rangle$, and hence the success probability of performing the $\logic{F_{d}}$ gate is constant. We refer the reader to Appendix~\ref{app:rot_gates_details} for a detailed derivation of Clifford gates with qudits in RSB.

To conclude this Section, we provide in Table~\ref{tab:gkpnoneencodings} the list of the CV physical operations that correspond to Clifford operations on GKP and RSB-encoded states. In the next Section, we will assess the  simulatability of these CV operations.

\begin{table}[htbp]
\caption{\label{tab:gkpnoneencodings}Clifford operations on computational basis encoded states in GKP and RSB, respectively. For GKP, we only consider Clifford operations for  even dimension $d$, due to our encoding method described in Sec.~\ref{sec:GKP_qubit_into_qudit}.
 In contrast, as will be described in Sec.~\ref{sec:qubit-into-qudit-RBC}, for RSB we can consider Clifford operations in any  dimension $d$, both even and odd. The parameter $\beta$ in the phase gate $S$ for RSB  is given by $\beta=0$ in even dimension, while,  $\beta=1$ in odd dimension. In Sec.~\ref{sec:our-CV-theorem} we prove that these operations, acting on Wigner negative encoded stabilizer GKP or RSB states and followed by measurement in the computational basis of the $d-$dimensional qudit, are  classically efficiently simulatable.}
\begingroup
\setlength{\tabcolsep}{2pt} 
\renewcommand{\arraystretch}{2.5} 
\begin{ruledtabular}
 \begin{tabular}{c c c c c}
  Operation
   & $\text{GKP}^{(2)}$
   & $\text{GKP}^{(d)}$
   & $\text{RSB}^{(2)}$
   & $\text{RSB}^{(d)}$                                                               \\
  \hline
  $\logic{Z}$
   & $e^{i\sqrt{\pi}\hat q}$
   & $e^{i\sqrt{\tfrac{2\pi}{d}}\hat q}$
   & $e^{i \frac{\pi}{N} \hat{n}}$
   & $e^{i\tfrac{2\pi}{dM}\hat{n}}$                                                   \\
  $\logic{F}$
   & $e^{i\frac{\pi}{4}(\hat p^2+\hat q^2)}$
   & $e^{i\frac{\pi}{4}(\hat p^2+\hat q^2)}$
   & Fig.~\ref{fig:teleportedFourier}
   & Fig.~\ref{fig:teleportedFourier}                                                 \\
  $\logic{S}$
   & $e^{i\hat q^2/2}$
   & $e^{i\hat q^2/2}$
   & $e^{i\tfrac{\pi}{2N^{2}}\hat{n}^2}$
   & $e^{i\tfrac{\pi}{d}\left(\frac{\hat{n}^2}{M^{2}}-\beta\frac{\hat{n}}{M}\right)}$ \\
  $\logic{C_Z}^{(k,l)}$
   & $e^{i\hat q_k \hat q_l}$
   & $e^{i\hat q_k \hat q_l}$
   & $e^{i\tfrac{\pi}{N^2}\hat{n}_{k}\hat{n}_{l}}$
   & $e^{i\tfrac{2\pi}{dM^2}\hat{n}_{k}\hat{n}_{l}}$                                  \\
  $\logic{X}$
   & $e^{-i\sqrt{{\pi}}\hat p}$
   & $e^{-i\sqrt{\tfrac{2\pi}{d}}\hat p}$
   & $\logic{F}^\dagger \logic{Z} \logic{F}$
   & $\logic{F}^\dagger \logic{Z} \logic{F}$                                          \\
  $\textsc{\logict{sum}}^{(k,l)}$
   & $e^{-i\hat{q}_k \hat{p}_l}$
   & $e^{-i\hat{q}_k \hat{p}_l}$
   & $\logic{F}^{(l)\dagger} \logic{C_{Z}} \logic{F}^{(l)}$
   & $\logic{F}^{(l)\dagger} \logic{C_{Z}} \logic{F}^{(l)}$                           \\
 \end{tabular}
\end{ruledtabular}
\endgroup
\end{table}

\section{Efficient simulation of continuous-variable architectures}
\label{sec:our-CV-theorem}
By virtue of the considerations that we developed in the previous Sections, we are now in the position to prove the main results of our paper. Specifically, we identify processes in CVs that can be simulated efficiently despite being characterized by Wigner-negative states and operations. First, we briefly review the central role of the Wigner function representation in the context of the simulatability of CV quantum circuits.

Systems of quantum CVs characterized by Gaussian Wigner functions have been extensively studied \cite{ferraro2005, weedbrook2012, adesso2014continuous}. In particular, it is well known that the process of performing arbitrary Gaussian operations and measurements (such as homodyne detection) on multipartite Gaussian states is efficiently simulatable~\cite{bartlett2002}. The same holds true for more general processes in which states, operations, and measurements can be described by non-negative Wigner functions ~\cite{mari2012, veitch2013}. This imposes the necessity of going beyond these settings, with the ultimate goal of achieving quantum computational advantage. In particular, this underpinned the introduction of the resource theory of Wigner negativity \cite{takagi2018, albarelli2018} and, more in general, stimulated an intensive effort in devising universal CV platforms in which quantum information can be hosted in Wigner-negative states \cite{yukawa2013, ulanov2017quantum, sabapathy2018states, fluhmann2018, brunelli2018unconditional, houhou2018unconditional, fluhmann2019, campagne-ibarcq2019}, and processed using Wigner-negative operations \cite{xiang2013hybrid, krastanov2015universal, albarelli2016nonlinearity, marek2018general, sabapathy2019production, gao2019entanglement, hillmann2020} and measurements \cite{walschaers2018tailoring, eaton2019non, su2019generation, ra2020non}.

Of specific interest for our purposes is the observation that, by supplying an otherwise simulatable sequence of Gaussian gates and homodyne measurement with input GKP states, the CV simulatability theorems of Refs. \cite{mari2012, veitch2013, rahimi-keshari2016} no longer hold. In fact, all pure encoded GKP states are characterized by a negative Wigner function, and so lie outside the non-negative Wigner criteria for simulatability. Even considering non-negative Wigner operations and measurement, we cannot use the CV theorems to assess efficient classical simulatability for the encoded architectures described in this Section. Note that, remarkably, it has been proven that fault-tolerant universal computation becomes possible for some specific Wigner-negative circuits of this type \cite{gottesman2001, baragiola2019}.

Given the potential importance of Wigner-negative circuits, it is therefore of relevance to scrutinize them in detail. To demonstrate, rather counter-intuitively, that large families of Wigner-negative circuits in CVs can be simulated efficiently, we consider DV systems embedded in CV ones via both GKP and RSB encoding. Therefore, given this interpretation, we can apply known results for efficiently simulating quantum circuits in DVs to investigate CV operations acting on the corresponding logical subspace of the embedded DV systems.

Crucially, we extend our results to include families of circuits beyond the immediate translation of DV theorems for qubits with bosonic encoding. The key step for this extension is the interpretation of encoded qubit states, or states of dimension $d_1$, as states belonging to a higher-dimensional system characterized by dimension $d_2$, thus defining a logical space within which we can accommodate a larger number of operations. In particular, we will interpret stabilizer GKP or RSB qubit states, at the input of a circuit, as stabilizer GKP or RSB qudit states, over which a larger class of operations (namely, Clifford operations corresponding to the enlarged logical space) can enact yet keeping the circuit simulatable.

\subsection{Architectures with translation-symmetric encodings}
\label{sec:gkp-qunits-to-qudits}
We start by analyzing what happens to initial translation-symmetric states (in particular, GKP qubits) when acted on with CV operations that do not preserve their code subspace, such as displacements that do not correspond to the encoded qubit displacements in Eq.~(\ref{eq:clifford-gkp0}).
Thus, we consider circuits of the form as in Fig.~\ref{fig:circuit}, where we initialize the inputs as GKP-encoded $\ket{\logic{0_2}}$ qubit states, but extend the range of operations to the encoded Clifford operations for $d$-dimensional systems. Before exploring these circuits in full generality, we first present an introductory example whereby Wigner negative circuits are efficiently simulatable.

\begin{figure}[htpb]
 \centering
 $$
  \Qcircuit @C=2.5em @R=1.2em {
  \lstick{\ket{\logic{0_2}}}   & \qw & \multigate{2}{\mathcal C_2=\{\logic{H},\logic{S},\textsc{\logict{cnot}}\}} & \qw & \measureD{\hat{q}} \\
  \lstick{\vdots\ \ \ }  & {/}\qw  & \ghost{\mathcal C_2=\{\logic{H},\logic{S},\textsc{\logict{cnot}}\}} &  {/}\qw & \vdots\\
  \lstick{\ket{\logic{0_2}}} & \qw & \ghost{\mathcal C_2=\{\logic{H},\logic{S},\textsc{\logict{cnot}}\}} &  \qw & \measureD{\hat{q}}
  }
 $$
 \caption{The input states are the logical qubit basis states encoded with GKP code in CVs. These states are operated on by any operation within the GKP-encoded Clifford group for qubits, described in Sec.~\ref{sec:GKP_qubit_into_qudit}. Homodyne measurements are performed in the position basis and correspond to Pauli $Z$ measurements in the qubit basis. The resulting circuit is efficiently simulatable even though the input states possess a negative Wigner function.}
 \label{fig:circuitqubit}
\end{figure}
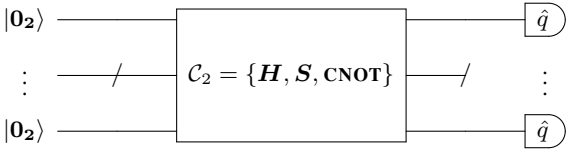

Fig.~\ref{fig:circuitqubit} explicitly shows the reduced form of Fig.~\ref{fig:circuit}, whereby we restrict to $d_1=2$.
The input states are the Wigner-negative GKP stabilizer states in Eq.~(\ref{eq:code-words-qubit}), which are operated on by the group of qubit Clifford operations described in Sec.~\ref{sec:gkp-qubits}, and measurements are performed by homodyne detection, which corresponds to encoded Pauli measurements~\cite{gottesman2001}.
The  correspondence between homodyne detection and GKP-encoded Pauli measurements can be understood in simple terms by inspecting the GKP wave-functions in position representation Eqs.~(\ref{eq:code-words-qubit}): it is clear that, in the ideal case of infinitely-squeezed codewords, measuring the position of the peaks, \textit{i.e.}, the observable $\hat q$, unambiguously determines whether the GKP qubit is in state $\ket{\logic{0_2}}$ or $\ket{\logic{1_2}}$, and corresponds therefore to a measurement in the computational basis.

Despite the Wigner negative input states, this circuit corresponds to a stabilizer circuit in DVs, which can be simulated efficiently on classical computers, as it is known from the Gottesman-Knill theorem~\cite{gottesman1999,nielsen2000}. However, there is no theorem in CV quantum computation to classify the process as simulatable, since the GKP-encoded $|\logic{0_2}\rangle$ is Wigner negative.

Lifting the restriction and allowing $d_2=2a^2$, as in Fig.~\ref{fig:circuit}, we can make a more powerful statement about the simulation of the evolution of the input GKP qubit states.
Expanding the code space in this way shows that more operations are classically efficiently simulatable than just the qubit Clifford encoded operations displayed in Fig.~\ref{fig:circuitqubit}. In particular, we can prove simulatability for displacements beyond the ones given in Eq.~(\ref{eq:clifford-gkp0}) for GKP qubits.

\subsubsection{GKP code: Interpretation of a lower-dimensional qudit state as a higher-dimensional qudit state}
\label{sec:GKP_qubit_into_qudit}
We have seen in Sec.~\ref{sec:qubits-qudits} that it is possible to embed a $d_1$-dimensional qudit into a qudit of dimension $d_2=d_1 a_1 a_2$ for some $a_1,a_2\in \mathbb N$. We are going to consider here the case where both $d_1$-dimensional and $d_2$-dimensional qudits are supported in a CV system via GKP encoding.

In particular, by considering $a=a_1=a_2$, $d_2=d_1 a^2$, we can observe that a GKP-encoded $d_1$-dimensional state $\ket{\logic{j_{d_1}}}$ also represents a $d_2$-dimensional GKP qudit state. Using Eq.~(\ref{eq:code-words-2}), it is possible to verify the following identity for non-normalized GKP-encoded codewords
\begin{align}
 \label{eq:GKP_d1_d2}
 \ket{\logic{j_{d_1}}}=\sum_{k=0}^{a-1}\ket{\logic{(aj+ad_1 k)_{d_2}}} ,
\end{align}
where the right-hand side denotes a superposition of the $(aj+ad_1 k)$-th states of the computational basis of a $d_2$-dimensional GKP qudit. The above identity can be proven by showing explicitly that both $d_1$- and $d_2$-dimensional logical states are described by the same wave-function in CVs. The $d_1$-dimensional state $\ket{\logic{j_{d_1}}}$ is given in the GKP encoding by
\begin{align}
 \label{eq:d1_GKP}
 \ket{\logic{j_{d_1}}} = \sum_{s\in\mathbb Z} \ket{\sqrt{\tfrac{2\pi}{d_1}}(j+d_1 s)}_{\hat q} ,
\end{align}
which coincides for $d_2=d_1 a^2$ with the $d_2$-dimensional GKP qudit state
\begin{align}
 \label{eq:d2_GKP}
 \sum_{k=0}^{a-1}\ket{\logic{(aj+ad_1 k)_{d_2}}} & = \sum_{k=0}^{a-1} \sum_{s\in\mathbb Z}
 \ket{\sqrt{\tfrac{2\pi}{d_2}}(aj+ad_1 k+d_2 s)}_{\hat q}\nonumber                         \\
                                                 & = \sum_{k=0}^{a-1} \sum_{s\in\mathbb Z}
 \ket{\sqrt{\tfrac{2\pi}{d_1}}[j+d_1 (k+ a s)]}_{\hat q}\nonumber                          \\
                                                 & = \sum_{s\in\mathbb Z}
 \ket{\sqrt{\tfrac{2\pi}{d_1}}(j+d_1 s)}_{\hat q} .
\end{align}

As an illustrative example, we consider $d_1 = 2$, and the GKP-encoded qubit state $\ket{\logic{0_2}}$. For $a = 2$, we could interpret the state as a $d_2=8$ dimensional GKP qudit
\begin{align}
 \label{eq:GKPtransform-example}
 \ket{\logic{0_2}} = \ket{\logic{0_8}}+\ket{\logic{4_8}}.
\end{align}
As seen in Sec.~\ref{subsec:stabiliser-formalism} and Appendix~\ref{app:generating_encoded_qubit}, the logical state $\ket{\logic{0_8}}+\ket{\logic{4_8}}$ defined in the higher dimension $d_2=8$ is a stabilizer state.
The same applies for the general case of Eq.~(\ref{eq:GKP_d1_d2}).

\begin{figure}[htpb]
 \centering
 \includegraphics[width=0.9\linewidth]{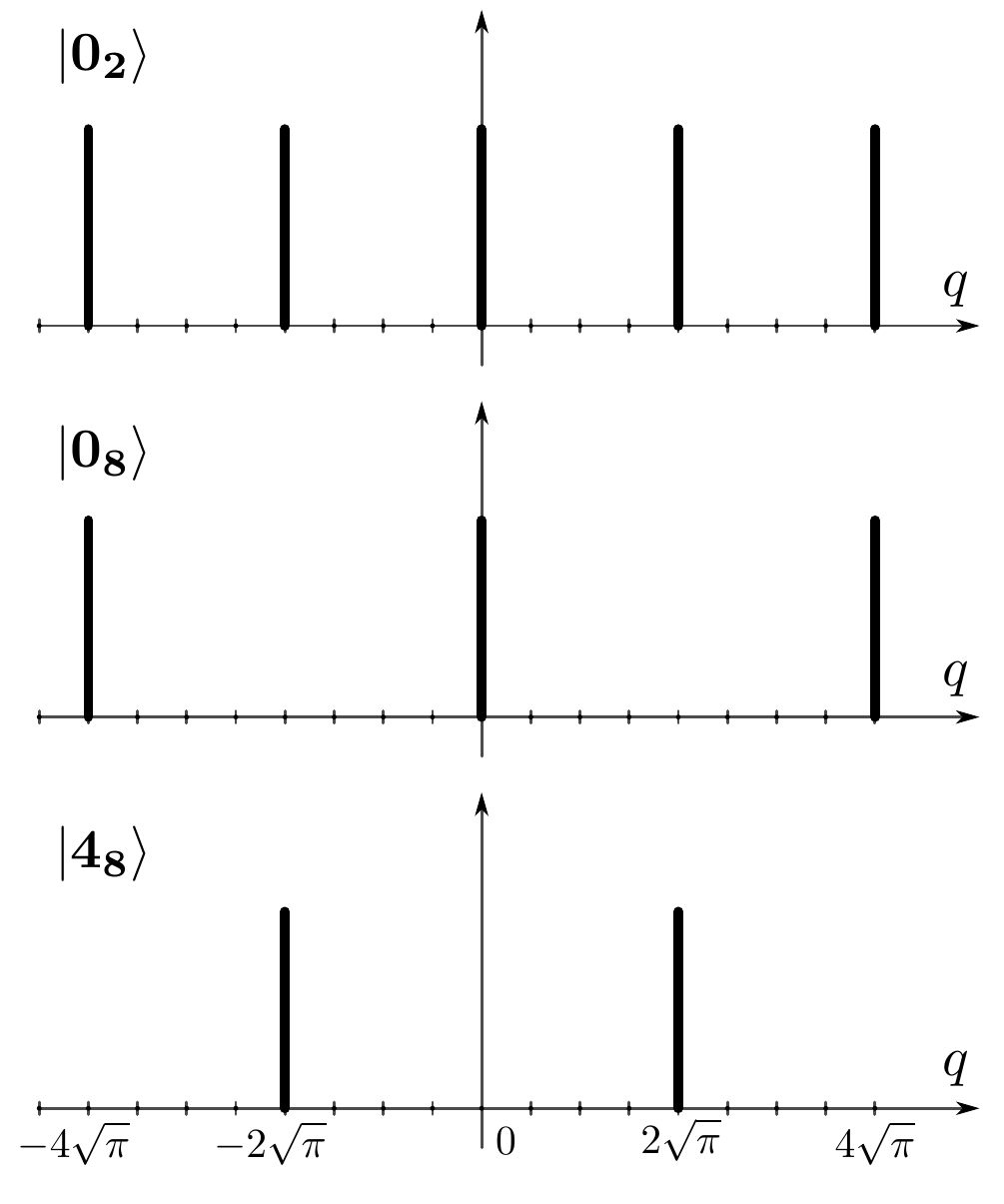}
 \caption{Wave-function representation of the non-normalized GKP-encoded states $\ket{\logic{0_2}}$, $\ket{\logic{0_8}}$, and $\ket{\logic{4_8}}$. We observe that the positions of the Dirac delta functions coincide, as expected from Eq.~(\ref{eq:GKPtransform-example}).}
 \label{fig:GKP02}
\end{figure}

We stress that, in comparison with the purely DV case of Eq.~(\ref{eq:qubit-qudit-DV}) and Eq.~(\ref{eq:qunit_qudit_DV}), here we are not merely mapping $d_1$-dimensional logical information into systems of higher dimension $d_2$.
Instead, in Eq.~(\ref{eq:GKP_d1_d2}) and Eq.~(\ref{eq:GKPtransform-example}), the \emph{wave-function} of $d_1$-dimensional GKP states (\textit{e.g.}, $\ket{\logic{0_2}}$) can be \emph{identified} as a higher-dimensional GKP states (\textit{e.g.},
$\ket{\logic{0_8}}+\ket{\logic{4_8}}$), as represented in Fig.~\ref{fig:GKP02}.
That is, both states are encoded logical states of dimension $d_1$ and $d_2$, respectively, as emphasized by the use of bold fonts in our notation.

Furthermore, the logical $\logic{X}$ and $\logic{Z}$ operators in $d_1$ dimensions can be interpreted as their $d_2$-dimensional equivalent operators applied $a$ times.  As we have seen in Sec.~\ref{sec:qubits-qudits}, in the case of symmetric encoding $d_2 = a^2 d_1$, the Clifford operation generators given by the Fourier transform, the phase gate and the $\textsc{\logict{sum}}$ in $d_2$-dimensions  also act as their equivalent Clifford operators in $d_1$-dimensions~\cite{farinholt2014}.

Notice that, for asymmetric encodings, such that $a_1\ne a_2$, we no longer expect to see this symmetry and so the higher-dimensional Fourier transform will no longer act as a Fourier transform on the qudits. These embeddings are non-symplectic in the sense of Ref.~\cite{farinholt2014}, and so there is no longer a guarantee that it is possible to implement the logical Clifford operations via operations in the $d_2$-dimensional space.
The following discussion will focus on symmetric encodings.

\subsubsection{GKP code: Extension of CV simulatable operations}
\label{sec:GKP_extended_CV}
As a consequence of the discussions above, for the circuits represented in Fig.~\ref{fig:circuit} that are composed of the operations in Eq.~(\ref{eq:GKP-operations}), and in particular by displacements $t\sqrt{2\pi/d_1}$ consisting of a rational $t$, we can choose a proper qudit dimension $d_2=d_1 a^2$, such that  $t= j/a$ for $j$ and $a$ integers.
Then, we interpret that physical displacement as logical displacements
\begin{align}
 \logic{Z_{d_2}}^j & = e^{i\hat q j\sqrt{\frac{2\pi}{d_2}}}= e^{i\hat q \frac{j}{a} \sqrt{\frac{2\pi}{d_1}}}, \nonumber \\
 \logic{X_{d_2}}^j & = e^{-i\hat p j\sqrt{{2\pi}/{d_2}}}= e^{-i\hat p \frac{j}{a} \sqrt{{2\pi}/{d_1}}},
\end{align}
where $j\in\mathbb Z_{d_2}$.
These are $d_2$-dimensional Clifford operations. We consider measurement through homodyne detection, which as said corresponds to Pauli measurement. Due to the theorems that we have presented in Sec.~\ref{sec:DV-theorems} for the efficient classical simulatability of Clifford circuits in dimension $d_2$, we can conclude therefore that the resulting architectures are classically efficiently simulatable in the sense of strong simulatability, despite the Wigner negativity of the input encoded GKP stabilizer states. The list of simulatable operations includes all operations that are interpreted as Clifford operations within the GKP qudit encoding in dimension $d_2=d_1 a^2$ for arbitrary integer $a$.
They include the Fourier transform, the entangling gate $e^{i\hat{q}_k\hat{q}_l}$, as well as the shear operation $e^{i\hat{q}_k^2/2}$. For convenience, we list again the simulatable operations given in Eq.~(\ref{eq:GKP-operations}),
\begin{align}
 \label{eq:GKP-operations-bis}
 \{e^{i\hat q_k^2/2},e^{-i\alpha\hat p_k},e^{i\alpha\hat q_k},e^{i\hat q_k\hat q_l},e^{\tfrac{i\pi}{4}(\hat p_k^2 + \hat q_k^2)}\}
\end{align}
where $\alpha = \sqrt{{2\pi}/{d_1}}/a$, for any $a$ integer. Note that in particular, it possible to displace $j$ times, which corresponds to the logical operation $\logic{X_{d_2}}^j$. The correspondence in terms of GKP-encoded qubit operations is summarized in Table~\ref{tab:gkpnoneencodings}.

Note that, although we have presented results of the simulatability of architectures with input GKP-encoded qudit states of any dimension $d_1$, one can focus on the particular example of input stabilizer qubit states.

\subsection{Architectures with rotation-symmetric encodings}
\label{sec:our-CV-theorem-RBC}
In analogy to what we have derived in Sec.~\ref{sec:gkp-qunits-to-qudits} for the GKP encoding, we aim now at encoding lower-dimensional qudits in higher-dimensional qudits within RSB codes. In contrast to the previous case, we can follow two methods described in Sec.~\ref{sec:qubit-into-qudit-RBC}. The first method allows for the encoding of $d_1$-dimensional states, for instance qubit computational basis states, in $d_2$-dimensional qudit states, similarly to the GKP case, with additional restrictions on the rotation symmetry.
On the other hand, the second method consists in the encoding of $|\logic{0_{d_1}}_{;N,\varphi}\rangle$ states in $|\logic{+_{d_2}}_{;M,\varphi}\rangle$ states of arbitrary dimension $d_2$.
In particular, we can choose $d_2$ to be odd prime, and therefore use additional results for classical simulation~\cite{gottesman1999a,dam2011,bravyi2016a,bravyi2016,bravyi2019,huang2019}.
In both cases, we need to consider additional conditions related with the rotation symmetry, as shown in Table~\ref{tab:stateencodings}, and we can extend the set of simulatable operations beyond those defined in the qubit logical space, as we address in Sec.~\ref{sec:RBC_extended_CV}.

\subsubsection{RSB code: Interpretation of a lower-dimensional qudit state as a higher-dimensional qudit state}
\label{sec:qubit-into-qudit-RBC}
In order to encode a lower-dimensional qudit in higher-dimensional qudits in RSB codes, we recall Eq.~(\ref{eq:qunit_qudit_DV}). For symmetric encoding of a qubit defined with $N$-fold rotation symmetry we have $d_2 = d_1 a^2$ with $a\in\mathbb{N}$, and the additional restriction $N\geq a$, since the qudit states should have $M$-rotational symmetry, with $M=N/a$. We recall the qudit computational basis states in rotational codes, given in Eq.~(\ref{eq:codewordrbc}).

We consider the cases in which the rotated primitive states are orthogonal, $\langle \varphi|\logic{Z_{d}}^{s}|\varphi\rangle = 0$, for $0<s<dN$, which implies that the normalization constants for all the computational basis states are identical and proportional to the rotation symmetry integer $N$ and the qudit dimension $d$, ${\mathcal{N}}_j = dN$.
Generally, the primitives in rotational codes are exactly orthogonal only in some appropriate limit~\cite{grimsmo2020}. In this case, the encoding is given by
\begin{align}
 \label{eq:rot_encod}
 |\logic{j_{d_1}}_{;N}\rangle = \frac{1}{\sqrt{a}} \sum_{t=0}^{a-1} |\logic{(aj + ad_{1}t)_{d_2}}_{;\frac{N}{a}}\rangle,
\end{align}
as derived in Appendix~\ref{app:encoding_details}. Similarly to the case of Eq.~(\ref{eq:GKP_d1_d2}), the right-hand side denotes a superposition of the $(aj+ad_1 k)$-th states of the computational basis of a $d_2$-dimensional RSB qudit.

\begin{figure*}[htpb]
 \centering
 \includegraphics[width=0.9\linewidth]{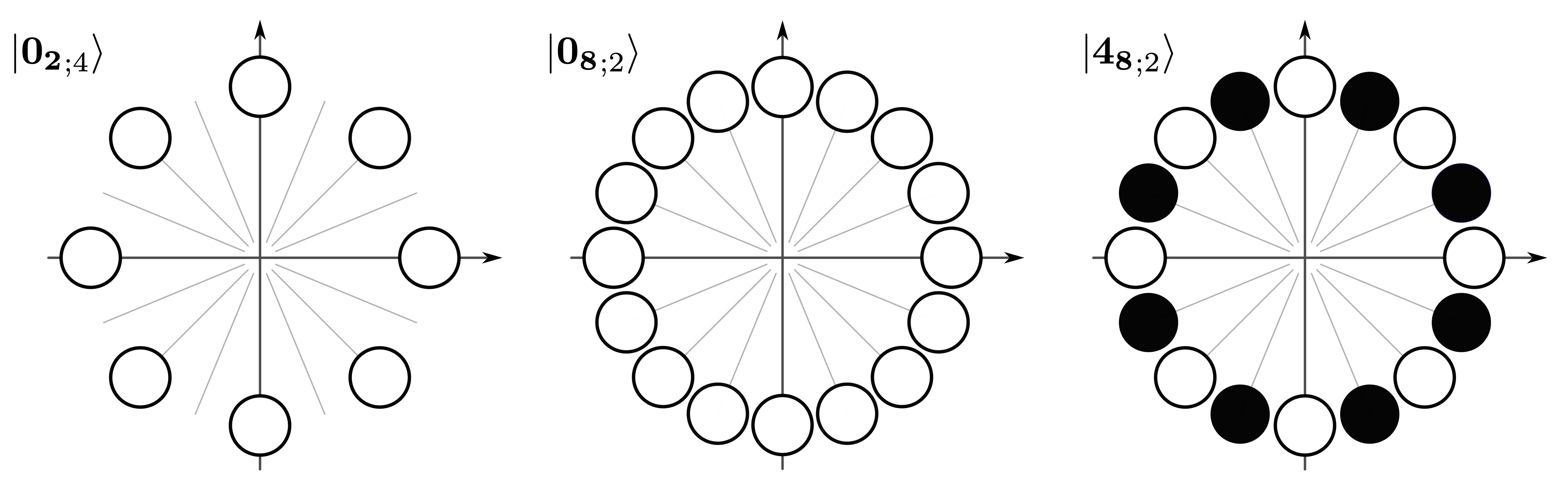}
 \caption{Phase-space graphical representation of the RSB-encoded states $\ket{\logic{0_2}_{;4}}$, $\ket{\logic{0_8}_{;2}}$, and $\ket{\logic{4_8}_{;2}}$. The circles represent primitive functions (\textit{e.g.}, coherent states), where the white and black colors refer to positive and negative prefactors, respectively. We observe that the combination of the rotated primitive functions $\ket{\varphi}$ of $\ket{\logic{0_8}_{;2}}$ and $\ket{\logic{4_8}_{;2}}$ as indicated in Eq.~(\ref{eq:RSBtransform-example}) recreates the desired state $\ket{\logic{0_2}_{;4}}$.}
 \label{fig:RSB02}
\end{figure*}

We consider the encoding of the qubit RSB state $\ket{\logic{0_2}_{;4}}$ as an example, where we have $d_1 = 2$, and $N=4$. For $a = 2$, we can interpret the state as a $d_2=8$ dimensional RSB qudit with $M=2$ rotation symmetry
\begin{align}
 \label{eq:RSBtransform-example}
 \ket{\logic{0_2}_{;4}} = \frac{1}{\sqrt{2}} \ket{\logic{0_8}_{;2}}+\ket{\logic{4_8}_{;2}},
\end{align}
as depicted in Fig.~\ref{fig:RSB02}.

An alternative encoding comes from the relation between the $|\logic{0_{d_1}}\rangle$ states in the $\logic{Z_{d_1}}$ basis, and the $|\logic{u^0_{d_2}}\rangle\equiv |\logic{+_{d_2}}\rangle$ states in the $\logic{X_{d_2}}$ basis, given in Eq.~(\ref{eq:Xd_codewords}).
Thus, for the case of orthogonal rotated primitive states, that is ${\mathcal{N}}_j=dM$  for $M$-folded rotational symmetry, we can substitute Eq.~(\ref{eq:codewordrbc}) in Eq.~(\ref{eq:Xd_codewords}) as shown in Appendix~\ref{app:encoding_details}, such that for dimension $d_2$ the $\logic{X_{d_2}}$ states are
\begin{align}
 \label{eq:Xd_code_prim}
 |\logic{u^k_{d_2}}_{;M,\varphi}\rangle = \frac{1}{\sqrt{M}} \sum_{\ell=0}^{M-1} e^{i \frac{2\pi}{{d_2}M} (\ell {d_2} - k)\hat{n}} |\varphi\rangle.
\end{align}
Hence, we notice that for $k=0$, the eigenstate $|\logic{u^0_{d_2}}_{;M,\varphi}\rangle  \equiv |\logic{+_{d_2}}_{;M,\varphi}\rangle$ has the same form for arbitrary dimension ${d_2}$.

Given the expression for $|\logic{+_{d_2}}_{;M,\varphi}\rangle$ for any dimension ${d_2}$ and rotation symmetry $M$, we observe that it corresponds to the state $|\logic{0_{d_1}}_{;N,\varphi}\rangle$ for dimension $d_1$ and rotation symmetry $N$ if $M=d_{1} N$.
Therefore, it is possible to encode the state $|\logic{0_{d_1}}_{;N,\varphi}\rangle$ for any dimension $d_1$  in any other dimension $d_2$, where it corresponds to the state $|\logic{+_{d_2}}_{;M,\varphi}\rangle$.
This in turn implies that any circuit initialized in the stabilizer state $|\logic{0_{d_1}}_{;N,\varphi}\rangle$ for a $d_1$-dimensional RSB code can be equivalently regarded as being initialized in the stabilizer state $|\logic{+_{d_2}}_{;M,\varphi}\rangle$ for a ${d_2}$-dimensional RSB code.

\subsubsection{RSB code: Extension of CV simulatable operations}
\label{sec:RBC_extended_CV}
In order to recognize the largest possible family of CV circuits in RSB codes as classically efficiently simulatable, we can take advantage of both the interpretations introduced above. A main difference between GKP Clifford operations and RSB Clifford operations is the fact that in RSB codes, in general, encoded Clifford operations can be Wigner negative. Hence, in this case, our results on simulatability also include families of circuits with Wigner-negative operations in addition to Wigner-negative initial states.

\begin{table}[htpb]
\caption{\label{tab:stateencodings}Interpretation of codewords defined in a logical space of dimension $d_1$ as codewords of a different logical space of dimension $d_2$, within the GKP and RSB codes. The wave-functions describing the states of the second and third column coincide if the conditions in the fourth column are fulfilled. For the GKP encoding, the details of the identification are given in Sec.~\ref{sec:GKP_qubit_into_qudit}.
 For RSB codes, it is possible to use two mappings with their respective conditions for interpreting the wave-functions as codewords of different logical spaces, as shown in Sec.~\ref{sec:qubit-into-qudit-RBC}.}
\begingroup
\setlength{\tabcolsep}{2pt} 
\renewcommand{\arraystretch}{2.5} 
\begin{ruledtabular}
 \begin{tabular}{c >{\centering\arraybackslash}p{0.22\linewidth} >{\centering\arraybackslash}p{0.45\linewidth} >{\centering\arraybackslash}p{0.18\linewidth}}

                                                                                          & Codespace of dimension $d_1$   & \hfil Codespace of \newline dimension $d_2$                & \hfil \newline Conditions        \\ \hline
  GKP                                                                                     & $|\logic{j_{d_1}}\rangle$      & $\sum\limits_{k=0}^{a-1}|\logic{(aj+ad_1 k)_{d_2}}\rangle$ & $d_2 = d_1 a^2$     \vspace{1mm} \\ \hline
  RSB                                                                                     & $|\logic{j_{d_1}}_{;N}\rangle$ &
  $\dfrac{1}{\sqrt{a}} \sum\limits_{t=0}^{a-1} |\logic{(aj + ad_{1}t)_{d_2}}_{;M}\rangle$ & $\begin{array}{c}
    d_2 = d_1 a^2 \\
    M={N}/{a}
   \end{array}$                                                                                                   \\ \hline
  RSB                                                                                     & $|\logic{0_{d_1}}_{;N}\rangle$ & $|\logic{+_{d_2}}_{;M}\rangle$                             & $\begin{array}{c}
    \forall d_1, d_2 \\
    M=d_1 N
   \end{array}$     \\
 \end{tabular}
\end{ruledtabular}
\endgroup
\end{table}

The first method relates RSB-encoded systems of dimension $d_1$ and rotation-symmetry of order $N$ with systems of dimension $d_2= d_1 a^2$ and rotation-symmetry of order $M=N/a$, with the additional restriction of $N\geq a$. This extends the number of Clifford operations from the ones corresponding to $d_1$-dimensional systems (in particular, for qubits with $d_1=2$) to the ones corresponding to $d_2$-dimensional systems. For example, we extend the number of simulatable rotations in phase space given by Eq.~(\ref{eq:zd_rot}). That is, instead of the set generated by
\begin{align}
 \label{eq:number_rot}
 \hat{R}_{d_1 N}=e^{i\tfrac{2\pi}{d_1 N}\hat{n}},
\end{align}
we can use the extended set of rotations generated by
\begin{align}
 \hat{R}_{d_2 M}
 = e^{i\tfrac{2\pi}{d_2 M}\hat{n}}
 =e^{i\tfrac{2\pi}{ad_1 N}\hat{n}},
\end{align}
where we have used that $d_2 = d_1 a^2$ and $M=N/a$, without affecting the simulatability of the corresponding circuits. We note that the number of rotations is higher despite the restriction on the order of the rotation-symmetry of the encoding.

With the second method, we can interpret the  $|\logic{0_{d_1}}_{;N,\varphi}\rangle$ computational states, for dimension $d_1$ and rotation symmetry $N$, as $|\logic{+_{d_2}}_{;M,\varphi}\rangle$ states for any dimension $d_2$ and $M$-order rotation symmetry, with $M=d_{1} N$.
As a result, we can decide to interpret the $d_1$-dimensional state as a state in a logical space of dimension $d_2$, with $d_2$ odd prime.
Indeed, in contrast to the GKP case and the first encoding method in RSB codes, this choice allows us to access additional algorithms for classical simulation of quantum circuits~\cite{gottesman1999a,dam2011,bravyi2016a,bravyi2016,bravyi2019,huang2019}. The number of operations that we can simulate is also extended in comparison to those defined in the $d_1$ logical space. As an illustrative example, we compare the number of rotations generated by the operator given in Eq.~(\ref{eq:number_rot}) with those generated by
\begin{align}
 \hat{R}_{d_2 M}= e^{i\tfrac{2\pi}{d_2 M}\hat{n}} = e^{i\tfrac{2\pi}{d_2 d_1 N}\hat{n}},
\end{align}
to conclude that we have enlarged the families of simulatable circuits in RSB codes. We stress that it is not possible to identify all codewords of the logical space of dimension $d_1$ as new codewords of the logical space of dimension $d_2$ with this second method, but only input $|\logic{0_{d_1}}_{;N}\rangle$ logical states.

\begin{table}[htpb]
\caption{\label{tab:RSBoperations}Extended set of Clifford operations for RSB codes for the two encoding methods described in Sec.~\ref{sec:qubit-into-qudit-RBC}. Here, $\beta=0$ for even dimension, and $\beta=1$ for odd dimension. In both cases, the set of physical simulatable operations increases with respect to the initial RSB code. Depending on the method, the new operations corresponding to the logical gates $\logic{S}$ and $\logic{C_Z}^{(k,l)}$ differ. Therefore, different families of extended Wigner negative circuits can be simulated classically depending on the method chosen.}
\begingroup
\setlength{\tabcolsep}{2pt} 
\renewcommand{\arraystretch}{2.5} 
\begin{ruledtabular}
 \begin{tabular}{>{\centering\arraybackslash}p{0.15\linewidth} >{\centering\arraybackslash}p{0.23\linewidth} >{\centering\arraybackslash}p{0.255\linewidth} >{\centering\arraybackslash}p{0.30\linewidth}}

  \hfil\newline \null\hfil \newline Operation & \hfil \textbf{Initial RSB:} \newline \null\hfil Dimension: $d_1$ \newline Order: $N$         & \hfil \textbf{Method 1:} \newline \null\hfil Dimension: $d_1 a^2$ \newline Order: $N/a$ & \hfil \textbf{Method 2:} \newline \null\hfil Dimension: $d_2$ \newline Order: $d_1 N$ \\ \hline
  $\logic{Z}$                                 & $e^{i\tfrac{2\pi}{d_1 N}\hat{n}}$                                                            & $e^{i\tfrac{2\pi}{d_1 a N}\hat{n}}$                                                     & $e^{i\tfrac{2\pi}{d_2 d_1 N}\hat{n}}$                                                 \\
  $\logic{S}$                                 & $e^{i\tfrac{\pi}{d_1}\left(\frac{\hat{n}^2}{N^{2}}-\beta\frac{\hat{n}}{N}\right)}$           &
  $e^{i\tfrac{\pi}{d_1}\left(\frac{\hat{n}^2}{N^{2}}-\beta\frac{\hat{n}}{a N}\right)}$
                                              & $e^{i\tfrac{\pi}{d_2 d_1}\left(\frac{\hat{n}^2}{{d_1} N^{2}}-\beta\frac{\hat{n}}{N}\right)}$                                                                                                                                                                                   \\
  $\logic{C_Z}^{(k,l)}$                       & $e^{i\tfrac{2\pi}{d_1 N^2}\hat{n}_{k}\hat{n}_{l}}$                                           & $e^{i\tfrac{2\pi}{d_1 N^2}\hat{n}_{k}\hat{n}_{l}}$                                      & $e^{i\tfrac{2\pi}{d_2 {d_1}^2 N^2}\hat{n}_{k}\hat{n}_{l}}$                            \\
 \end{tabular}
\end{ruledtabular}
\endgroup

\end{table}

The discussion above can be summarized by stating that the circuits represented in Fig.~\ref{fig:circuitrsb}, composed by operations interpreted as qudit encoded Clifford operations, followed by phase measurements~\cite{holevo2011,helstrom1969}, are classically efficiently simulatable
due to the theorems of Sec.~\ref{sec:DV-theorems}.
Beyond the rotations addressed above, the physical operations associated with Clifford operations include a finite set of cross-Kerr and self-Kerr interactions, while Pauli measurements correspond to phase measurements (PM). In summary, this yields the simulatable families of circuits given previously in Eq.~(\ref{eq:RSB-operations-1}) and in Eq.~(\ref{eq:RSB-operations-2}). With the first embedding method, we could simulate the evolution of RSB-encoded input states in the computational basis, $|\logic{j_{d_1}}_{;N}\rangle$ with $j=0,\dots,d_1 -1$, when followed by a circuit generated by the interactions
\begin{align}
 \label{eq:RSB-operations-bis-1}
 \{e^{i\tfrac{2\pi}{ad_1 N}\hat{n}},e^{i\tfrac{\pi}{d_1}\left(\frac{\hat{n}^2}{N^{2}}-\beta\frac{\hat{n}}{a N}\right)},e^{i\tfrac{2\pi}{d_1 N^2}\hat{n}_{k}\hat{n}_{l}},\textrm{PM}\},
\end{align}
with $a$ a fixed natural number, and where $\beta$ takes values $\beta=0$ for even $d_1 a^2$, and $\beta=1$ for odd.

The second method  allows us to consider a different set of interactions as efficiently simulatable, for RSB input states $|\logic{0_{d_1}}_{;N}\rangle$.
The families of circuits that can be computed classically are those generated by
\begin{align}
 \label{eq:RSB-operations-bis-2}
 \{e^{i\tfrac{2\pi}{d_2 d_1 N}\hat{n}}, e^{i\tfrac{\pi}{d_2 d_1}\left(\frac{\hat{n}^2}{{d_1} N^{2}}-\beta\frac{\hat{n}}{N}\right)},e^{i\tfrac{2\pi}{d_2 {d_1}^2 N^2}\hat{n}_{k}\hat{n}_{l}},\textrm{PM}\}.
\end{align}
The parameter $\beta$ takes values $\beta=0$ for even dimension $d_2$, and $\beta=1$ for odd dimension $d_2$, for a chosen $d_2$. We remark here that classical simulatability is established despite the fact that not only RSB stabilizer codewords are represented by highly negative (even unbounded) Wigner functions, but also the operators in Eqs.~(\ref{eq:RSB-operations-bis-1}) and (\ref{eq:RSB-operations-bis-2}) involve non-Gaussian Wigner-negative transformations.

In particular  consider qubit input states, \textit{i.e.}, $d_1=2$. In that example, a circuit initialized in logical qubit computational states $|\logic{0_2}_{;N}\rangle$ and $|\logic{1_2}_{;N}\rangle$ within an RSB code, followed by operations that take the system outside the qubit logical space, can be simulated for particular families of interactions which can be interpreted as logical in higher-dimensional systems. In Table~\ref{tab:gkpnoneencodings}, we compare the physical CV operations associated with the qubit and the qudit subspaces, respectively.

\section{Implications for quantum superiority and fault-tolerance}
\label{sec:implications}

In this Section we briefly discuss the implications of our results on the classical simulatability of CV architectures with bosonic codes for architectures aiming at quantum advantage demonstrations with CV, as well as for quantum error-correction, where bosonic codes are an essential building block.

\subsection{Sampling models and quantum superiority}
Tailored continuous-variable architectures, such as boson samplers~\cite{wang2019}, have been outlined as candidates to prove quantum superiority when sampling from specific output probability distributions, analogously to qubit-based random circuit samplers~\cite{arute2019}. In particular, specific CV architectures with input non-Gaussian states such as photon-added and photon-subtracted squeezed states, Gaussian measurements and homodyne detection have been proved to provide quantum advantage, in the sense that the output probability distribution of the measurement results cannot be sampled efficiently by a classical computer~\cite{chabaud2017}. Our results show that for the goal of constructing one such computational model exhibiting quantum superiority in sampling, it is not enough to consider, for example, GKP states in computational basis encoded states, in combination with the operations expressed by Eq.~(\ref{eq:GKP-operations-bis}). One way of constructing one such model would be instead to consider GKP-encoded magic states as input and encoded Clifford operations, followed by homodyne detection. The computational hardness of the resulting architecture then follows from results available for qubit-based circuits~\cite{yoganathan2019}.

\subsection{Implications for fault-tolerance: Simulatability of error-correction circuits}

Consider a quantum computation encoded in GKP states.
At a certain step of the computation, the encoded data qubit encoded might have accumulated noise. For instance, this could be due to finite squeezing in the nodes of the supporting cluster state, if the calculation is performed by measurement-based quantum computation~\cite{gu2009}.

For a general data qubit, the GKP error-correction gadget circuit introduced in Ref.~\cite{menicucci2014} has the role of restoring the quantum information in the data qubit. After a GKP error correction is performed, noise in the data qubit is reduced. Due to the threshold theorem, this procedure (possibly concatenated with discrete-variable types of codes) is efficient, i.e., it is possible to restore exponential precision in the computation result with a polynomial number of quantum gates and ancillary GKP qubits.

\begin{figure}[h]
$$
\Qcircuit @C=1.0em @R=.7em {
\lstick{\ket{\psi}}  & \qw & \ctrl{1} & \qw & \qw &  \gate{X(-s \, \text{mod}[\sqrt{\pi}])} & \qw \\
\lstick{\ket{\logic{0_2}}}  & \qw & \control\qw  & \qw & \measureD{\hat p}& \control \cw \cwx & \lstick{s}
}
$$
\caption{\label{figECGadget}Procedure to correct for errors in the $\hat q$ quadrature. $\ket\psi$ is the data qubit and $\ket{0_L}$ is a GKP state. After measurement on the second mode the result $s$ is used to shift the first mode back.}
\end{figure}
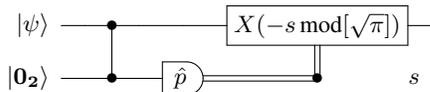

Many of the elements in the error-correction gadget described above are of the type considered in this work, and shown to be simulatable: the ancillary GKP state is in the $\ket{\logic{0_2}}$ logical state, entanglement of the data and ancillary qubit is achieved by a $C_Z$-type of interaction, and the syndrome measurement is performed by homodyne detection (see Fig.~\ref{figECGadget}). However, on the one hand, the displacement to perform depends on the measurement result, and it might not be of the type considered in our work. On the other hand, the state of the encoded data qubit $\ket{\psi}$ might correspond to a non-stabilizer state in general. Therefore, for encoded general quantum computations our work does not imply that error-corrected circuits become simulatable, and therefore is not in contradiction with previous results on general fault-tolerant quantum computation.

For some specific instances, the specific portion of circuit dedicated to the GKP error-correcting gadget exemplified by Fig.~\ref{figECGadget} might become classically simulatable. This occurs e.g. when the data qubits represent stabilizer states, and the measurement result is one of the simulatable displacements expressed in Eq.~(\ref{eq:GKP-operations-bis}). Note that this also happens with standard DV circuits, when one happens to perform a Clifford operation on stabilizer states, which might occasionally occur during a computation. In general, the classification of whether the circuit is simulatable or not depends on the specific combination of encoded data qubit, displacement, and adaptivity Ref.~\cite{koh2017further}.

\section{Outlook and conclusions}
\label{sec:outlook-conclusions}
In this work we have uncovered a previously unnoticed link between two mature areas of research within quantum information: on the one hand, the research on classical simulatability of DV-based quantum circuits and, on the other hand, the research on bosonic codes for fault tolerance of CV-based architectures. This allowed us to develop a framework for assessing the classical simulatability of CV architectures that are based on translation-symmetric and rotation-symmetric bosonic codes and are composed of the operations in Table~\ref{tab:gkpnoneencodings}, for which existing criteria could not be used. Our method relies on the interpretation of encoded lower-dimensional qudit states as encoded higher-dimensional qudit states given by the same bosonic code, and on considering existing algorithms that allow for the classical simulation of DV high-dimensional systems. We stress that this interpretation happens not only at the level of the logical quantum information, but also as the identification of the bosonic wave-functions of qubit and qudit states, living in Hilbert spaces with different dimensions, but sharing the same CV support through GKP or RSB codes.

Our proof of efficient classical simulatability for the architectures outlined above is non-trivial, due to the unsuitability of previously existing algorithms based on the CV Wigner function~\cite{mari2012,veitch2012,pashayan2015}.
In particular, in Ref.~\cite{pashayan2015} it was shown that the classical simulation cost of a general quantum computing architecture is exponential in a combination of the Wigner negativity of the input state, of the evolution operator, and of the measurement of the circuit considered, using a Monte Carlo algorithm that can handle the negative sign of the Wigner function~\cite{Note10}. While the technical derivations of Ref.~\cite{pashayan2015} dwell with discrete-variable systems, the same conclusions would hold in case one would consider CV systems, and use a discretization of the CV Wigner function as introduced by ~\cite{veitch2013}. The use of that algorithm for the simulation of our architectures would, therefore, result in an exponential simulation time in the size of the circuit.
Instead, our work points to the existence of classical algorithms, based on the generalizations of the Gottesman-Knill theorem to an arbitrary dimension $d$~\cite{gottesman1999a,hostens2005,beaudrap2013}, which, used in combination with a mapping of CV architectures to DVs through bosonic codes, allows for the efficient simulation of the CV architectures presented above.

Our results broaden the class of simulatable circuits in CV systems by including processes with negative Wigner states and operations. In particular, they corroborate the fact that the presence of Wigner negativity is not sufficient for computational speedup. This situation resembles the situation encountered when one considers the interplay between the potential speedup of pure-state quantum circuits and the degree of entanglement within them: on the one hand, pure-state computation with a low degree of entanglement is efficiently simulatable~\cite{jozsa2003role}; on the other hand, again appealing to the Gottesman-Knill theorem~\cite{gottesman1999}, there exist large families of quantum circuits that manifest a high degree of entanglement yet they are efficiently simulatable.
Analogously, for the case of CV Wigner negativity, circuits that display no or low degree of Wigner negativity are efficiently simulatable but, as we have proven here, there exist large families of quantum circuits that can be efficiently simulated even though they host a large, or even unbounded, degree of Wigner negativity.

We leave as an open question whether an arbitrary quantum state in CVs described by a negative Wigner function with certain symmetries can be interpreted as an encoded DV quantum state, such that we could apply our results to identify simulatable operations.
This question connects to the following one, beyond the use of bosonic encoding: which input states with Wigner negativity yield an output probability distribution that is hard-to-sample for a classical computer, when the rest of the circuit is Gaussian, and which do not? It is known, for instance, that some Gaussian circuits are made hard to sample by input single-photon states, or photon added and photon subtracted squeezed states~\cite{chabaud2017}.  In this work, we have shown that this is not the case for the Gaussian circuits that we have outlined, with input GKP stabilizer states.

Finally, it is still an open question whether arbitrary CV displacements --- namely, characterized by a \textit{continuous} parameter, rather than discrete displacements as in Eq.~(\ref{eq:GKP-operations}) --- acting on GKP-encoded states yield an architecture that is classically efficiently simulatable (and, similarly, arbitrary CV rotations for RSB-encoded states). However, this question is more meaningful to address in the realistic case of finitely squeezed states. We leave the analysis of this non-ideal case for future work, towards further refinements of the boundary
separating computationally useful and useless CV quantum architectures, be they CV NISQs~\cite{hillmann2020} or encoded CV circuits.

Note added: recently, we became aware of a work where the expectation values of the quadrature operators is shown to be classically efficiently computable for some classes of circuits with negative Wigner function~\cite{budiyono2020}.

\section{Acknowledgements}
G. F. and C. C. acknowledge support from the VR (Swedish Research Council) Grant QuACVA. G. F., L. G.-\'{A}., and C. C. acknowledge support from the Wallenberg Center for Quantum Technology (WACQT).

\appendix
\section{Wigner function formalism}
\label{app:wigner}

For the sake of completeness, we summarize here the notions of relevance for our purposes regarding the Wigner function description of states, operators, and measurements.

Following the notation introduced in the main text, we consider $n$ bosonic modes and collect the canonical position and momentum operators as a vector $\hat{\vec{r}}=\left(\hat{q}_1, \hat{p}_1 , \dots ,  \hat{q}_n , \hat{p}_n \right)$. We also introduce the corresponding vector of classical phase space variables $\vec{r} = \left(q_1, p_1 , \dots ,  q_n , p_n \right)$.
Assuming units such that $\hbar = 1$, the canonical bosonic commutation relations are $\left[ \hat{x}_i, \hat{p}_j\right] = i \delta_{ij}$, or more compactly $\left[ \hat{\vec{r}},\hat{\vec{r}}^\mathsf{T} \right] = i \Omega$, where the canonical symplectic form $\Omega$ is defined as
\begin{equation}
 \Omega = \bigoplus_{i=1}^N \omega, \qquad \omega = \begin{pmatrix}
  0  & 1 \\
  -1 & 0
 \end{pmatrix}.
\end{equation}
The Wigner function of a generic operator $\hat{O}$ acting on the Hilbert space of $n$ bosonic modes is defined as
\begin{equation}
 \mathcal{W}[\hat{O}](\vec{r})= \frac{1}{ \left( 2 \pi \right) ^{2 n} } \int_{\mathbb{R}^{2n}} \mathrm{d} \vec{v} \, e^{i \vec{v}^\mathsf{T} \Omega \vec{r}} \Tr \left[ \hat{O} e^{- i \vec{v}^\mathsf{T} \Omega \hat{\vec{r}}} \right]\,.
\end{equation}
A detailed analysis of the properties of the Wigner function representation can be found for example in Refs.~\cite{ferraro2005,serafini2017}.

Let us consider first the case in which the operator $\hat{O}$ represents a quantum state. By virtue of Hudson's theorem~\cite{hudson1974}, it is known that the only pure states having a non-negative Wigner function are Gaussian states, namely states whose Wigner function is Gaussian. All the pure states used as basis for the code-spaces considered in this work have a non-Gaussian Wigner function, that assumes negative, possibly unbounded, values for some region of the phase space. For simplicity, with a slight abuse of language, we say that such states have a \textit{negative} Wigner function.

Consider now the Wigner function representation of unitary operations such as the ones that constitute gates in the quantum circuits of Figs.~\ref{fig:circuit} and \ref{fig:circuitrsb}. In CV systems, unitary operations preserving the Gaussian character of the Wigner function --- and therefore the non-negativity of the Wigner function for pure input states --- have generators that are at most quadratic functions of the position and momentum operators, as for example the unitary operators in Eq.~(\ref{eq:GKP-operations}).
Such operations admit, via the Jamiolkowski isomorphism, a description in terms of non-negative Gaussian Wigner functions~\cite{mari2012}. On the other hand, unitaries generated by functions of higher order in the position and momentum operators have negative Wigner functions, as for example the  unitary operator $e^{i\tfrac{2\pi}{d_1 N^2}\hat{n}_{k}\hat{n}_{l}}$ in Eq.~(\ref{eq:RSB-operations-1}).

Finally, a description of measurements in term of Wigner functions can also be given. A fundamental property of the Wigner function is that the trace of two operators can be expressed as an integral in phase space
\begin{equation}
 \label{eq:tr_rule_ph_space}
 \Tr[\hat{O}_1 \hat{O}_2] = (2 \pi)^n \int_{\mathbb{R}^{2n}} \mathrm{d} \vec{r} \mathcal{W}[ \hat{O}_1] ( \vec{r}) \mathcal{W}[\hat{O}_2] (\vec{r}),
\end{equation}
in particular this can be used to express the Born rule
\begin{equation}\label{eq:wig_born_rule}
 p(\vec{a}|\rho)= (2 \pi)^n \int_{\mathbb{R}^{2n}} \mathrm{d} \vec{r} \mathcal{W}[ \rho ] ( \vec{r}) \mathcal{W}[\Pi_{\vec{a}}] (\vec{r}),
\end{equation}
where $\int_\Gamma \mathrm{d} \mu(\vec{a}) \Pi_{\vec{a}} = \mathbb{I}$ is a generic positive-operator valued measurement (POVM). The integral measure $\mu(\vec{a})$ on the outcome space $\Gamma$ is generic --- \textit{e.g.}, the Lebesgue measure for general-dyne measurements or the counting measure for photon-counting measurements \cite{serafini2017}.
When the Wigner function of the POVM effect $\mathcal{W} \left[ \Pi_{\vec{a}} \right] (\vec{r})$ is a Gaussian function we refer to these as Gaussian measurements. Of relevance for our purposes, notice that the case of homodyne measurements HM is described by non-negative Wigner functions (in particular, the Wigner function of the effect is a Dirac delta in the limit of infinite resolution). On the other hand, phase measurements PM have negative Wigner functions.

\section{Clifford operations in symmetric encodings}

\label{app:stabilizer_preservation}
In the following we give an example of an operation that preserves its stabilizer status in both dimension $d_1$ and $d_2=d_1 a^2$.
The Clifford operation generators given by the Fourier transform, the phase gate and the $\textsc{SUM}$ in $d_2$-dimension also act as their equivalent Clifford operators in $d_1$-dimensions~\cite{farinholt2014}.
As an example, consider the logical state of a qubit $d_1=2$ encoded within a system of dimension $d_2=2\cdot 3^2$. The state is given by
\begin{align}
 \label{example-encoding-qubit-18}
 \ket{\logic{0_2}} = \frac{1}{\sqrt 3}\left(\ket{0_{18}} + \ket{6_{18}} + \ket{12_{18}} \right) ,
\end{align}
which is stabilized in the $d_2=18$-dimensional qudit space by the generating set  $\langle X^6_{18}, Z^6_{18}\rangle$. Applying the quantum Fourier transform on the qudit will correspond to applying the Hadamard gate on the logical qubit,
\begin{align}
   & F_{18}\ket{\logic{0_2}} =                                                                                                                                    \nonumber         \\
 = & \left(\frac{1}{\sqrt{18}}\sum_{j,k =0}^{17} \omega^{jk}_{18}\ket k \bra j\right)\frac{1}{\sqrt 3}\left(\ket 0_{18} + \ket 6_{18} + \ket{12}_{18} \right)\nonumber              \\
 = & \frac{1}{3\sqrt {6}}\left(\sum_{k=0}^{17} \omega^{0k}_{18}\ket k_{18}+\sum_{k=0}^{17} \omega^{6k}_{18}\ket k_{18}+\sum_{k=0}^{17} \omega^{12k}_{18}\ket k_{18}\right)\nonumber \\
 = & \frac{1}{\sqrt{6}}\left(\ket{0}_{18}+\ket{3}_{18}+\ket{6}_{18}+\dots+\ket{15}_{18}\right) \nonumber                                                                            \\
 = & \frac{1}{\sqrt 2} \left(\ket{\logic{0_2}}+\ket{\logic{1_2}}\right)=\ket{\logic{+_2}} ,
\end{align}
with $\omega_{18}=e^{i2\pi/18}$. Note that for asymmetric encodings, such that $a_1\ne a_2$, this is no longer true. The logical Clifford operators in $d_1$-dimension can not necessarily be defined with respect to their $d_2$-dimensional parent space Clifford operators.

\section{Generating an encoded stabilizer qudit state with Clifford operations}
\label{app:generating_encoded_qubit}

\begin{figure}[htp]
 \centering
 $$
  \Qcircuit @C=1.5em @R=1.2em {
  \lstick{\ket{0}_{d_2}^{(1)}}   & \gate{F_{d_2}} & \multigate{1}{\left(\textsc{sum}_{d_2}^{(1,2)}\right)^{a}} & \qw & \gate{X_{d_2}^{-t}} & \rstick{\ket{\logic{0_2}}} \qw  \\
  \lstick{\ket{0}_{d_2}^{(2)}} & \qw & \ghost{\left(\textsc{sum}_{d_2}^{(1,2)}\right)^{a}}  & \measureD{\mathcal M_{Z_{d_2}}} & \control \cw \cwx & \lstick{at}
  }
 $$
 \caption{A circuit to generate the $\ket{\logic{0_2}}$ state from any general $\ket{0}_{d_2}$ with dimension $d_2=2a^2$. After initializing both qudits in the $\ket{0}_{d_2}$ state, a Fourier transform is applied to the first qudit.
  This is followed by $a$ repetitions of the $\textsc{sum}$ gate. Finally, the second qudit is measured in the $Z$ basis and the result $t$ is used to perform the $X^{-t}$ feedback operation. The result is a $\ket{\logic{0_2}}$ state.}
 \label{fig:circuit_generate_encoded_qubit}
\end{figure}
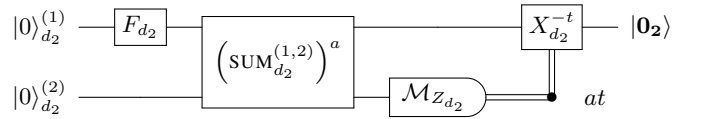

In this Appendix, we aim at showing that the state $\ket{\logic{0_2}} = \frac{1}{\sqrt {a}} \sum_{j=0}^{a-1} \ket{2aj}_{d_2}$ encoded in  dimension $d_2=2a^2$ can be generated from the state $\ket{0}_{d_2}$ using only Clifford operations, and hence is a stabilizer state. The Clifford circuit used for this purpose requires an additional qudit in a stabilizer state, which is measured and then discarded.
The state generated by the circuit in  Fig.~\ref{fig:circuit_generate_encoded_qubit} before measurement is given by
\begin{align}
  & \left(\textsc{sum}_{d_2}^{(1,2)}\right)^a F_{d_2}^{(1)}\ket{0}_{d_2} \ket{0}_{d_2} =                    \nonumber            \\
  & \quad =\frac{1}{\sqrt{d_2}}\left(\textsc{sum}_{d_2}^{(1,2)}\right)^a \sum_{k=0}^{d_2-1}\ket{k}_{d_2} \ket{0}_{d_2} \nonumber \\
  & \quad=\frac{1}{\sqrt{d_2}}\sum^{d_2-1}_{k=0}\ket{k}_{d_2}^{(1)}\ket{\left(ak\mod d_2\right)}_{d_2}^{(2)}.
\end{align}
The state of the second qudit, encoded in dimension $d_2$, is expressed as a sum over states of the form $\ket{\left(ak\mod d_2\right)}_{d_2}$ for some $k$. We could equally  write this as $\ket{a\left(k\mod 2a\right)}_{d_2}$, and so in the following we will write the second qudit state with respect to a dummy variable $t=k\mod{2a}$.

Therefore the state of the entire system before measurement can equally be written in the form
\begin{align}
  & \frac{1}{\sqrt{d_2}}\sum^{a-1}_{j=0} \sum^{2a-1}_{k=0}\ket{\left(k\mod 2a\right) + 2a j}_{d_2}^{(1)}\ket{a \left(k\mod 2a\right)}_{d_2}^{(2)}
 \nonumber                                                                                                                                        \\
  & = \frac{1}{\sqrt{d_2}}\sum^{a-1}_{j=0}\sum^{2a-1}_{t=0}\ket{(2aj+t)}_{d_2}^{(1)}\ket{at}_{d_2}^{(2)}.
\end{align}
A Clifford feedback operation is then applied to the first qudit proportional to the measured value of the second qudit. The purpose of this feedback operation is to subtract the $t$ dummy variable from the first qudit. The second qudit can then be discarded and the target state is achieved.

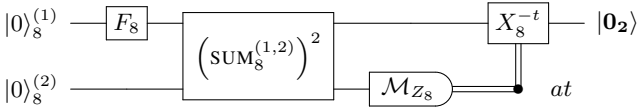
\begin{figure}[htpb]
 \centering
 $$
  \Qcircuit @C=1.5em @R=1.2em {
  \lstick{\ket{0}_{8}^{(1)}}   & \gate{F_{8}} & \multigate{1}{\left(\textsc{sum}_{8}^{(1,2)}\right)^{2}} & \qw & \gate{X_{8}^{-t}} & \rstick{\ket{\logic{0_2}}} \qw \\
  \lstick{\ket{0}_{8}^{(2)}} & \qw & \ghost{\left(\textsc{sum}_{8}^{(1,2)}\right)^{2}}  & \measureD{\mathcal M_{Z_8}}& \control \cw \cwx & \lstick{at}
  }
 $$
 \caption{This circuit is a specific example of the circuit given in Fig.~\ref{fig:circuit_generate_encoded_qubit} with dimension $d_2=8$ and $a=2$. The circuit generates the $\ket{\logic{0_2}}$ state from the $\ket{0}_{8}$ state. After initializing both qudits in the $\ket{0}_{8}$ state, a Fourier transform is applied to the first qudit.
  This is followed by $2$ repetitions of the $\textsc{SUM}$ gate. Finally, the second qudit is measured in the $Z_8$ basis and the result $t$ is used to perform the $X^{-t}$ feedback operation. The result is a $\ket{\logic{0_2}}$ state.}
 \label{fig:circuit_generate_encoded_qubit_example}
\end{figure}

As an example, consider the case that $a=2$. We want to show that the state  $\ket{\logic{0_2}} = \frac{1}{\sqrt {2}}\left(\ket{0}_8+\ket{4}_8\right)$ encoded in dimension $8$ can be generated from $\ket{0}_8$ with Clifford operations only.
This can be done using the circuit given in Fig.~\ref{fig:circuit_generate_encoded_qubit_example}.
The final state of the second qudit will be measured, with equal probability, as $t=0,1,2,3$. The final state of the first qudit will be
\begin{align}
 \frac{1}{\sqrt{2}}\left(\ket{(0+t)}_8 +\ket{(4+t)}_8 \right) .
\end{align}
By applying a feedback operation $X^{-t}$ dependent on the measured value, the target state $\ket{\logic{0_2}}$ will be recovered.

\section{Phase gate in the GKP code}
\label{app:sum-gate}
In this Appendix, we aim at providing an explicit derivation of the exponential form of the phase gate presented in Eq.~(\ref{eq:clifford-gkp3-bis}).

For qubits, the phase gate is defined as
\begin{align}
 S = \begin{pmatrix}
  1 & 0 \\ 0 & i
 \end{pmatrix}.
\end{align}

For GKP-encoded qudits, the phase gate is defined as the operator which acts on the displacement operators as~\cite{gottesman2001}
\begin{alignat}{3}
 \logic{X_d} & \to \eta_d \logic{X_d} \logic{Z_d} &  & \implies e^{i\sqrt{\frac{2\pi}{d}}\hat p}   &  & \to \eta_d e^{i\sqrt{\frac{2\pi}{d}}\hat p} e^{-i\sqrt{\frac{2\pi}{d}}\hat q} \nonumber \\
 \logic{Z_d} & \to \logic{Z_d}                    &  & \implies  e^{-i\sqrt{\frac{2\pi}{d}}\hat q} &  & \to  e^{-i\sqrt{\frac{2\pi}{d}}\hat q}
\end{alignat}
where $\eta_d=\omega_{D}\omega_{2d}^{-1}$.

This operation can be realized by a symplectic transformation which acts on the continuous variables as~\cite{gottesman2001}
\begin{align}
 \label{eq:qtoq}
 \hat q & \to \hat q  \nonumber     \\
 \hat p & \to \hat p - \hat q + c ,
\end{align}
where $c=0$ for even $d$ and $c=\sqrt{\frac{\pi}{2d}}$ for odd $d$. We aim at verifying here that the operation which performs this transformation can be expressed as Eq.~(\ref{eq:clifford-gkp3-bis}), that we report here for convenience,
\begin{align}
 \logic{S_d} = e^{i(\hat q^2-2c\hat q)/2} .
\end{align}
This operation commutes with $\hat q$ and hence satisfies Eq.~(\ref{eq:qtoq}). Since $\hat q$ commutes with itself, the operation can also be rewritten as
\begin{align}
 \logic{S_d} = e^{i\hat q^2/2}e^{-ic\hat q} .
\end{align}
Applying this to the $\hat p$ operator and recognizing that $e^{ic\hat q}$ is a displacement by $c$ in momentum space, we find that
\begin{align}
 \logic{S_d} \hat p \logic{S_d}^\dagger & = e^{i\hat q^2/2}\left(\hat p+c\right) e^{-i\hat q^2/2} \nonumber \\
                                        & = e^{i\hat q^2/2}\hat pe^{-i\hat q^2/2} + c .
\end{align}
Using the Baker-Campbell-Hausdorff lemma
\begin{align}
 e^{\hat A}\hat B e^{-\hat A}=\hat B +[\hat A,\hat B] + \frac{1}{2!}[\hat A,[\hat A,\hat B]] +\dots ,
\end{align}
we see that
\begin{align}
 e^{i\hat q^2/2}\hat p e^{-i\hat q^2/2}= & \hat p +\frac{i}{2}[\hat q^2,\hat p] + \frac{i^2}{2!\cdot2^2}[\hat q^2,[\hat q^2,\hat p]] +\dots \nonumber \\
 =                                       & \hat p +\frac{i}{2}(2\hat q i) + \frac{i^2}{2!\cdot2^2}[\hat q^2,(2\hat q i)] +\dots \nonumber             \\
 =                                       & \hat p - \hat q .
\end{align}
We can, therefore, conclude that
\begin{align}
 \logic{S_d}\hat p \logic{S_d}^\dagger   = \hat p - \hat q + c .
\end{align}
As an example, consider the action of $S_2$ on the GKP-encoded $\ket{\logic{0_2}}$ and $\ket{\logic{1_2}}$ states,
\begin{align}
 \logic{S_2} \ket{\logic{0_2}} = & \sum_{s\in\mathbb{Z}}e^{i\hat q^2/2}\ket{2s\sqrt\pi}_{\hat q}\nonumber \\
 =                               & \sum_{s\in\mathbb{Z}}e^{i2s^2\pi}\ket{2s\sqrt\pi}_{\hat q}\nonumber    \\
 =                               & \ket{\logic{0_2}} ,
\end{align}
and
\begin{align}
 \logic{S_2} \ket{\logic{1_2}} = & \sum_{s\in\mathbb{Z}}e^{i\hat q^2/2}\ket{(2s+1)\sqrt\pi}_{\hat q}\nonumber       \\
 =                               & \sum_{s\in\mathbb{Z}}e^{i(2s+1)^2\pi/2}\ket{(2s+1)\sqrt\pi}_{\hat q}\nonumber    \\
 =                               & \sum_{s\in\mathbb{Z}}e^{i(2s^2+2s+1/2)\pi}\ket{(2s+1)\sqrt\pi}_{\hat q}\nonumber \\
 =                               & e^{i\pi/2}\ket{\logic{1_2}} = i \ket{\logic{1_2}}.
\end{align}

\section{Universal quantum computation in bosonic codes}
\label{app:t_gates}
Universal quantum computation in the GKP and RSB schemes with encoded qubits can be achieved by adding the magic gate, $T=|0\rangle \langle 0|+ e^{i\pi/4}|1\rangle \langle 1|$, to the Clifford set of operations.

Besides the circuits for gate-teleportation of the $T$ operation~\cite{nielsen2000} one can define the encoded operation implementing a logic $\logic{T}$ gate in both GKP and RSB encodings.
In both cases this operation is not uniquely defined.

For GKP-encoded qubits, it is possible to implement the $\logic{T}$ gate with the unitary operator~\cite{gottesman2001}
\begin{align}
 \label{eq:Tgate_GKP1}
 \logic{T} = e^{{i\frac{\pi}{4}}\left[
    2\left(\frac{\hat{q}}{\alpha}\right)^3+\left(\frac{\hat{q}}{\alpha}\right)^2-2\left(\frac{\hat{q}}{\alpha}\right)\right]},
\end{align}
where for GKP symmetric states, $\alpha = \sqrt{\pi}$ is half of the spacing that determines the GKP translational symmetry, or alternatively, by
\begin{align}
 \label{eq:Tgate_GKP2}
 \logic{T} = e^{i\frac{\pi}{4}\left(\frac{\hat{q}}{\alpha}\right)^{4}}.
\end{align}

For RSB-encoded qubits with $N$-fold symmetry, the $\logic{T}$ gate is given by the unitary operation
\begin{align}
 \label{eq:Tgate_RBC1}
 \logic{T} = e^{{i\frac{\pi}{4}}\left[
    2\left(\frac{\hat{n}}{N}\right)^3+\left(\frac{\hat{n}}{N}\right)^2-2\left(\frac{\hat{n}}{N}\right)\right]},
\end{align}
or alternatively, by~\cite{grimsmo2020}
\begin{align}
 \label{eq:Tgate_RBC2}
 \logic{T} = e^{i\frac{\pi}{4N^{4}}\hat{n}^{4}}.
\end{align}

\section{Fock-space structure of qudits in RSB codes}
\label{app:fock_qudit}
The Fock-space structure of the qudits can be analyzed by the action of $\logic{Z_d}$ on arbitrary states expanded in the Fock basis, $|\logic{\Psi_d}\rangle=\sum_{n=0}^{\infty} a_{n}|n\rangle$. We observe that the general form of the computational basis states for a code of dimension $d$ with $M$-fold symmetry is $|\logic{j_d}_{;M}\rangle = \sum_{s=0}^{\infty} a_{j,s}|jM + sdM\rangle$, as it follows from the following relation:
\begin{align}
 \label{eq:Fock_jcodes}
 \logic{Z_d} \sum_{s=0}^{\infty} a_{j,s}|jM + sdM\rangle = e^{i\frac{2\pi}{d}j} \sum_{s=0}^{\infty} a_{j,s}|jM + sdM\rangle.
\end{align}
The coefficients $a_{j,s}$ can be written in terms of the primitive function characteristic of the encoding $|\varphi\rangle$, as we relate both representations with the identity between a train of Kronecker deltas with period $M$ and a summation of complex exponentials
\begin{align}
 \label{eq:deltas-formula}
 \frac{1}{M}\sum_{m=0}^{M-1} e^{\pm i \frac{2\pi mn}{M}} = \sum_{k= -\infty}^\infty \delta_{n,kM} ,
\end{align}
and the projectors onto different sets of Fock states for a general state $\sum_{n=0}^{\infty} c_n |n\rangle$~\cite{albert2014},
\begin{align}
 \hat{\Pi}_{dM}^{jM} \sum_{n=0}^{\infty} c_n |n\rangle= & \sum_{s=0}^\infty |sdM+jM\rangle \langle sdM+jM| \sum_{n=0}^{\infty} c_n |n\rangle                                                     \nonumber \\
 =                                                      & \sum_{n=0}^{\infty} \sum_{s=0}^{\infty} \delta_{n-jM,sdM} c_n |n\rangle                                                            \nonumber     \\
 =                                                      & \sum_{n=0}^{\infty} \frac{1}{dM}\sum_{m=0}^{dM-1} e^{i \frac{2\pi m(n-jM)}{dM}} c_{n} |n\rangle                                       \nonumber  \\
 =                                                      & \frac{1}{dM}\sum_{m=0}^{dM-1} \left(e^{-i\frac{2\pi j}{d}} e^{i \frac{2\pi m\hat{n}}{dM}} \right)^m \sum_{n=0}^{\infty} c_{n}|n\rangle \nonumber \\
 =                                                      & \frac{1}{dM}\sum_{m=0}^{dM-1} \left(e^{-i\frac{2\pi j}{d}} \logic{Z_d} \right)^m \sum_{n=0}^{\infty} |n\rangle,
\end{align}
which allow us to write the logical encoded states as
\begin{align}
 |\logic{j_d}_{;M,\varphi}\rangle = \frac{dM}{\sqrt{{\mathcal{N}}_j}} \hat{\Pi}_{dM}^{jM}|\varphi\rangle.
\end{align}
Thus, the coefficients of Eq.~(\ref{eq:Fock_jcodes}) are proportional to the projection amplitude of the primitive state to different Fock states, $a_{j,s}=dM \langle sdM+jM|\varphi\rangle / \sqrt{\mathcal{N}_j}$.

\section{Qudit operations in RSB codes}
\label{app:rot_gates_details}
We analyze the action of qudit Clifford operations defined in Sec.~\ref{sec:rot_codes} for discrete RSB codes, when applied onto the codewords in the Fock-space basis.

The phase gate for qudits is defined in Eq.~(\ref{eq:phase-gate-clifford}). That is, $\logic{S_{d}} = \sum_{j=0}^{d-1} e^{i\frac{\pi}{d}j^2}\ket{\logic{j_d}}\bra{\logic{j_d}}$ for even dimensions,
and $\logic{S_{d}} = \sum_{j=0}^{d-1} e^{i\frac{\pi}{d}(j^2 - j)}\ket{\logic{j_d}}\bra{\logic{j_d}}$ for odd-dimensional qudits.
For even dimensions, the RSB-encoded phase gate is the self-Kerr interaction shown in Eq.~(\ref{eq:phase_rot_qudit}) with $\beta=0$, and therefore, its action when applied onto the logical encoded states defined in the Fock space as $|\logic{j_d}_{;M}\rangle =\sum_{s=0}^{\infty} a_{j,s}|jM + sdM\rangle$ is
\begin{align}
 \logic{S_d} |\logic{j_d}_{;M}\rangle & = e^{i\tfrac{2\pi}{2dM^{2}}\hat{n}^2} \sum_{s=0}^{\infty} a_{j,s}|jM + sdM\rangle                     \nonumber  \\
                                      & = \sum_{s=0}^{\infty} e^{i\tfrac{2\pi}{2dM^{2}}(jM + sdM)^2}  a_{j,s}|jM + sdM\rangle                  \nonumber \\
                                      & = \sum_{s=0}^{\infty} e^{i\pi s^{2} d} e^{i 2\pi sj} e^{i\tfrac{2\pi}{2d}j^2} a_{j,s}|jM + sdM\rangle \nonumber  \\
                                      & = \sum_{s=0}^{\infty} e^{i\tfrac{2\pi}{2d}j^2} a_{j,s}|jM + sdM\rangle,
\end{align}
where we have taken into account that $d$ is an even number larger than $2$ if we encode a qubit in higher dimensions.

The phase gate for qudits of odd dimension is another self-Kerr interaction, defined in Eq.~(\ref{eq:phase_rot_qudit}) for $\beta=1$. Likewise, its action when applied onto the computational basis states defined in the Fock space is given by
\begin{align}
 \logic{S_d} |\logic{j_d}_{;M}\rangle & = e^{i\tfrac{2\pi}{2d} \left(\frac{\hat{n}^2}{M^{2}}-\frac{\hat{n}}{M}\right)} \sum_{s=0}^{\infty} a_{j,s}|jM + sdM\rangle                     \nonumber  \\
                                      & = \sum_{s=0}^{\infty} e^{i\tfrac{2\pi}{2d}\left(\frac{(jM + sdM)^2}{M^{2}}-\frac{jM + sdM}{M}\right)}  a_{j,s}|jM + sdM\rangle                  \nonumber \\
                                      & = \sum_{s=0}^{\infty} e^{i\pi (s^{2} d - s)} e^{i 2\pi sj} e^{i\tfrac{2\pi}{2d}(j^2-j)} a_{j,s}|jM + sdM\rangle \nonumber                                 \\
                                      & = \sum_{s=0}^{\infty} e^{i\tfrac{2\pi}{2d}(j^2-j)} a_{j,s}|jM + sdM\rangle,
\end{align}
where we have taken into account that for $d$ an odd number, and that $s^{2} d - s$ is an even number larger than $2$ for all any odd or even $s$.

The entangling gate $\logic{C_Z}$ between two qudits with $N$ and $M$ rotation symmetry is described by the interaction shown in Eq.~(\ref{eq:controlZ_rot_qudit}). Then, its action when applied onto the codewords defined in the Fock space is
\begin{align}
  & \logic{C_Z}^{({k}_{N},{s}_{M})} |\logic{\ell_d}_{;N}\rangle^{(k)} |\logic{j_d}_{;M}\rangle^{(s)} = e^{i\tfrac{2\pi}{dNM}\hat{n}_{k}\hat{n}_{s}} \nonumber \\
  & \qquad \times \sum_{m=0}^{\infty} a_{m}|\ell N + mdN \rangle^{(k)} \sum_{t=0}^{\infty} a_{t}|jM + tdM \rangle^{(s)} \nonumber                             \\
  & = \sum_{m,t=0}^{\infty} e^{i\tfrac{2\pi}{d}(mtd^2 + jmd + \ell td + \ell j)}  \nonumber                                                                   \\
  & \qquad \times a_{m}a_{t}|\ell N + mdN \rangle^{(k)} |jM + tdM \rangle^{(s)}\nonumber                                                                      \\
  & = e^{i\tfrac{2\pi}{d} \ell j} |\logic{\ell_d}_{;N}\rangle^{(k)} |\logic{j_d}_{;M}\rangle^{(s)}.
\end{align}

The circuit shown in Fig.~\ref{fig:teleportedFourier} describes the teleported $\logic{F_d}$ gate. The probability of measuring the eigenstate $|\logic{u^{0}_d}{;M,\varphi}\rangle$, 
and hence of performing the gate successfully, is constant.
The conditional probability describing the post-selection is well-defined for any $\logic{X_d}$ eigenstate $|\logic{u^{n}_d}{;M,\varphi}\rangle$, where $n=0,1,\dots, d-1$,
\begin{align}
  & \langle\logic{\Psi_f}|\left(|\logic{u^{n}_d}{;M,\varphi}\rangle\langle \logic{u^{n}_d}{;M,\varphi}|\sum_{\ell=0}^{d-1}|\logic{\ell_d}\rangle\langle\logic{\ell_d}|\right)|\logic{\Psi_f}\rangle =                                                                                                                                                           \nonumber \\
  & =\frac{1}{d} \left(\sum_{j,k=0}^{d-1} \alpha^{*}_k e^{-i\tfrac{2\pi}{d}jk} \frac{1}{\sqrt{d}}e^{-i\tfrac{2\pi}{d}nk}\delta_{j,\ell}\right) \nonumber                                                                                                                                                                                                                  \\
  & \qquad \times \left(\sum_{p,m=0}^{d-1} \alpha_p e^{i\tfrac{2\pi}{d}pm} \frac{1}{\sqrt{d}}e^{i\tfrac{2\pi}{d}np}\delta_{\ell,m}\right)                               \nonumber                                                                                                                                                                                         \\
  & =\frac{1}{d^2} \sum_{j,k,p=0}^{d-1} \alpha^{*}_k \alpha_p e^{-i\tfrac{2\pi}{d}j(k-p)} e^{-i\tfrac{2\pi}{d}n(k-p)}                                                                                                                                                                              \nonumber                                                              \\
  & =\frac{1}{d} \sum_{k,p=0}^{d-1} \sum_{\ell=-\infty}^{\infty} \delta_{k-p,\ell d} \alpha^{*}_k \alpha_p e^{-i\tfrac{2\pi}{d}n(k-p)}                                                                                                                                                              \nonumber                                                             \\
  & =\frac{1}{d} \sum_{k=0}^{d-1} \alpha^{*}_k \alpha_k =\frac{1}{d},
\end{align}
where we have used the identity of Eq.~(\ref{eq:deltas-formula}).

It is then obvious to show that the probability to measure any eigenstate $|\logic{u^{n}_d}{;M,\varphi}\rangle$ is constant, $1/d$, and it does not depend on the input state $|\logic{\psi_d}\rangle$.

\section{Qudits embedded in higher-dimensional systems in RSB codes}
\label{app:encoding_details}
In the case of orthogonal rotated primitives, the encoding of qudits in higher-dimensional states shown in Eq.~(\ref{eq:rot_encod}) is exact. We derive explicitly,
\begin{align}
  & \frac{1}{\sqrt{a}} \sum_{t=0}^{a-1} |\logic{(aj + ad_{1}t)_{d_2}}_{;\frac{N}{a}}\rangle \nonumber                                                                                                                                                                              \\
  & = \frac{1}{\sqrt{a}} \sum_{t=0}^{a-1} \frac{1}{\sqrt{d_{2} N/a}}
 \sum_{m=0}^{d_{2} N/a -1} e^{-i\frac{2\pi}{d_{2}}(aj + ad_{1}t)m} e^{i\frac{2\pi}{d_{2}N/a}m\hat{n}}|\varphi\rangle \nonumber                                                                                                                                                     \\
  & =  \frac{1}{\sqrt{d_{2} N}} \sum_{m=0}^{d_{1} aN -1} \sum_{t=0}^{a-1} e^{-i\frac{2\pi}{d_{1}a}(j + d_{1}t)m} e^{i\frac{2\pi}{d_{1} aN}m\hat{n}}|\varphi\rangle                                                                                                       \nonumber \\
  & = \frac{1}{\sqrt{d_{2} N}} \sum_{m=0}^{d_{1} aN -1} e^{-i\frac{2\pi}{d_{1}a}jm}
 \left( a \sum_{k= -\infty}^{\infty} \delta_{m,ka} \right) e^{i\frac{2\pi}{d_{1} aN}m\hat{n}}|\varphi\rangle \nonumber                                                                                                                                                             \\
  & = \frac{1}{\sqrt{d_{2} N}} \sum_{k=0}^{d_{1} N -1} e^{-i\frac{2\pi}{d_{1}}jk} e^{i\frac{2\pi}{d_{1} N}k\hat{n}}|\varphi\rangle,
\end{align}
where we taken into account Eq.~(\ref{eq:deltas-formula}), and $0\leq m \leq d_1 a N -1$ for the limits of the summation over $k$.

The alternative encoding, in which the $|\logic{0_{d_1}}_{;N,\varphi}\rangle$ state corresponds to the $|\logic{+_{d_2}}_{;M,\varphi}\rangle \equiv |\logic{u^{0}_{d_2}}_{;M,\varphi}\rangle$ state if $M=d_{1} N$, can be demonstrated explicitly by considering the $\logic{X_{d_2}}$ eigenstates
\begin{align}
 |\logic{u^{k}_{d_2}}_{;M,\varphi}\rangle & = \frac{1}{{d_2}\sqrt{M}} \sum_{j=0}^{{d_2}-1} e^{-i\frac{2\pi}{{d_2}}kj} \sum_{m=0}^{{d_2}M-1} e^{-i \frac{2 \pi}{{d_2}} jm} e^{i \frac{2\pi}{{d_2}M} m\hat{n}} |\varphi\rangle \nonumber \\
                                          & = \frac{1}{{d_2}\sqrt{M}} \sum_{m=0}^{{d_2}M-1} \sum_{j=0}^{{d_2}-1} e^{-i\frac{2\pi}{{d_2}}j(k+m)} e^{i \frac{2\pi}{{d_2}M} m\hat{n}} |\varphi\rangle                       \nonumber     \\
                                          & = \frac{1}{\sqrt{M}} \sum_{m=0}^{{d_2}M-1} \sum_{\ell=-\infty}^{\infty} \delta_{k+m,\ell {d_2}} e^{i \frac{2\pi}{{d_2}M} m\hat{n}} |\varphi\rangle                   \nonumber             \\
                                          & = \frac{1}{\sqrt{M}} \sum_{\ell=0}^{M-1} e^{i \frac{2\pi}{{d_2}M} (\ell {d_2} - k)\hat{n}} |\varphi\rangle.
\end{align}
We have consider the identity of Eq.~(\ref{eq:deltas-formula}), and $0\leq m \leq d_2 M -1$ for the limits of the summation over $\ell$.

\bibliographystyle{apsrev4-2}
\bibliography{./paper_stab.bib}

\end{document}